\documentclass[useAMS,usenatbib]{mn2e}
\usepackage{graphicx,epsfig,amssymb,amsmath,layout,verbatim,rotating,calc,mathrs
fs,natbib}
\usepackage{graphics}
\usepackage{rotate}
\usepackage{float} 
\usepackage{txfonts}
\def\O{$\Omega$}
\input wasyfont

\newcommand{\dgr}{$^{\circ}$}
\newcommand{\ej}{E$_{J}$ }

\def\emp{\mbox{\raisebox{.4mm}
{\tiny{$\rangle$\hspace{-.2mm}$\bigcirc$\hspace{-.15mm}$\langle$}}}}

\title[The phase space of peanuts]{The phase-space of boxy-peanut and
{\sf X}-shaped bulges in galaxies\\ I. Properties of non-periodic orbits}
\author[P.A.~Patsis  \& M.~Katsanikas]
{P.A.~Patsis,$^{1,2}$\thanks{patsis@academyofathens.gr (PAP)}
M.~Katsanikas,$^1$\thanks{mkatsan@academyofathens.gr (MK)}\\
$^1$Research Center for Astronomy, Academy of Athens, Soranou Efessiou 4, GR-115
27, Athens, Greece\\
$^2$Excellence Cluster Universe, Boltzmannstr. 2, D-85748, Garching, Germany
}
\date{Accepted ..........Received .............;in original form ..........}

\begin{document}
\maketitle

\label{firstpage}
 
\begin{abstract}
The investigation of the phase-space properties of structures encountered in a
dynamical system is essential for understanding their formation and
enhancement. In the present paper we explore the phase space in energy intervals
where we have orbits that act as building blocks for boxy-peanut (b/p) and
``{\sf X}-shaped'' structures in rotating potentials of galactic type. We
underline the significance of the rotational tori around the 3D families x1v1
and x1v1$^{\prime}$ that have been bifurcated from the planar x1 family. These
tori play a multiple role: (i) They belong to quasi-periodic orbits that
reinforce the local density. (ii) They act as obstacles for the diffusion of
chaotic orbits and (iii) they attract a large number of chaotic orbits that
become sticky to them. There are also bifurcations of unstable families (x1v2,
x1v2$^{\prime}$). Their unstable asymptotic curves wind around the x1v1 and
x1v1$^{\prime}$ tori generating orbits with hybrid morphologies between that of
x1v1 and x1v2. In addition, a new family of multiplicity 2, called x1mul2, is
found to be important for the peanut construction. This family produces
stickiness phenomena in the critical area of the radial and vertical inner
Lindblad resonances (ILRs) and reinforces b/p bulges. Our work shows also that
there are peanut-supporting orbits before the vertical ILR. Non-periodic orbits
associated with the x1 family secure this contribution as well as the support of
b/p structures at several other energy intervals. Non-linear phenomena
associated with complex instability of single and double multiplicity families
of periodic orbits show that these structures are not interrupted in regions
where such orbits prevail. Depending on the main mechanism behind their
formation, boxy bulges exhibit different morphological features. Finally our
analysis indicates that ``{\sf X}'' features shaped by orbits in the
neighbourhood of x1v1 and x1v1$^{\prime}$ periodic orbits are pronounced only in
side-on or nearly end-on views of the bar.
\end{abstract}


\section{Introduction}
\label{sec:intro}
The skeleton of the boxy and peanut-like (b/p) structures observed in edge-on
disc galaxies is considered to be the families of periodic orbits (hereafter
p.o.), which constitute the x1-tree. These are the families bifurcating from the
planar x1 family \citep{gco81} which lies on the equatorial plane, at the $n:1$
vertical resonances \citep{spa02a,spa02b,psa02}. Despite the fact that the
periodic orbits determine their environment in the phase-space, the main body of
the structures are non-periodic orbits. It is known that even in 2D Hamiltonian
systems quasi-periodic orbits trapped around stable periodic orbits do not
necessarily support a structure of similar morphology. A morphology similar to
the one of the p.o. is secured mainly by quasi-periodic orbits on invariant
curves close to it. Contrarily, quasi-periodic orbits on invariant curves at the
outer parts of stability islands may support more complicated structures. For
example quasi-periodic orbits trapped around elliptical x1 orbits in 2D Ferrers
bars can support an ansae type bar, if orbits closer to the borders of the x1
stability islands are populated \citep{p05}. This independence between the
morphologies of the p.o. and of the structures around them is expected to be
more pronounced in 3D systems. Also the extent and the significance of the
phenomenon of stickiness \citep{ch10,bma} in building a 3D bar and a b/p
bulge has not been investigated up to now. 

In a recent series of papers \citep{kp11, kpc11, kpp11, kpc13} we have presented
the dynamical fingerprints of nonperiodic orbits close to p.o. in 3D rotating
Hamiltonian potentials of galactic type. For this purpose we used the
colour-rotation method proposed by \citet{pz94} in order to visualise the 4D
spaces of sections. In this way we have associated typical structures in the 4D
space with the neighbourhood of stable, as well a with simple, double and
complex unstable periodic orbits \citep{br69,h75,cm85}. Studies of other
Hamiltonian systems of galactic type \citep{ch13}, or even studies of maps
\citep{langetal, retal13, zkp13}, in which the 4D space is visualised, indicate
that the main structures depend basically on the corresponding kind of
instability, or  stability, of the p.o. In the present paper we take advantage
of the colour-rotation technique in order to investigate the phase space of the
b/p and {\sf X}-shaped structures in our model. This method has been proven to
be especially efficient in estimating the role of sticky orbits \citep{kpc13}. 

Referring to b/p-shaped bulges we mean boxy structures, the morphology of which
can be schematically described either with the symbol ``\emp'' or with the
symbol \begin{large}``$\infty$''\end{large}.  \citet{betal06} and \citet{ar06}
characterise these two qualitatively different profiles of edge-on disc galaxies
as ``OX'' and ``CX'' respectively. In many cases wings of an ``{\sf X}'' feature
delineate the peanut shape of the bulge, so the terms ``OX'' and ``CX'' refer to
an off-centred or centred ``{\sf X}'' respectively.

We assume that the {\sf X} feature traces the extrema of the b/p bulge. Let us
assume that it reaches a maximum radius along the major axis $R_{max}$ and a
maximum height vertical to the equatorial plane $z_{max}$. As we concluded in
\cite{psa02} the b/p bulge is part of the 3D bar, which is in agreement with the
results of studies with different approaches to the same problem \citep[see
e.g.][]{ldp00, ba05, betal06}. Observationally, the estimation of the extent of
the bar in edge-on galaxies is usually based on identifying ``bumps'' in the
surface brightness profiles in cuts along or parallel to their major axes as
indicated originally by \cite{wh84}. The qualitative and quantitative analysis
of such profiles allowed \cite{ldp00} to estimate the ratio of the length of the
bar to the length of the b/p distortion to be about $2.7\pm0.3$. \cite{ar06}
(his appendix C) presented the unsharp-masked profiles of a sample of 25 edge-on
galaxies with b/p bulges, which reveal a clear {\sf X} feature  \citep[see
also][]{betal06}. By measuring approximately the dimensions of the {\sf X}
features in the unsharp-masked images of this sample, one can empirically
estimate roughly a variation of the $r_{b/p}=R_{max}/z_{max}$ ratio, to be
$0.8\lessapprox r_{b/p} \lessapprox 1.5$. We estimated a mode value close to
$r_{b/p}=1.4$. This quantity depends primarily on the orientation of the bar
with respect to the line of sight and is expected to be small for bars viewed
nearly end-on and large for bars viewed nearly side-on, since $z_{max}$ will not
vary substantially. This is important information for deciding which orbits
populate a b/p structure. The goal of our study is to identify orbits which
could reinforce these structures and reproduce the observed ratios of their
geometrical dimensions.

The study of the body of the orbital content of b/p bulges can provide valuable
information about the properties of the stellar component of standard galactic
bars similar to the bar of the Milky Way \citep{szw11, ls12, wg13, eskoetal14}.
Non-periodic orbits reflect the dynamical mechanism that determines the volume
and the regions of the phase space, which they occupy. Moreover they provide
kinematic imprints that can be compared with observed quantities
\citep{kvkf04, wetal11,vzhetal13} or with results of $N$-body simulations
\citep{am02, ba05, so13, eskoetal14, qetal14}. This comparison can lead in
favouring or
excluding dynamical mechanisms and help us understand the internal structure of
the peanuts. The first step towards this comparison is to register all types of
orbital behaviour, regular and chaotic, that leads in profiles observed in
edge-on disc galaxies.

The paper is structured as follows:
The model used and a brief review on the stability of p.o. in 3D Hamiltonian
systems is given in section~\ref{sec:model}. In section~\ref{sec:start} we
discuss the origin of the b/p bulge, investigating the lowest energy orbits that
could support a b/p feature.  We present the dynamical
phenomena in the neighbourhood of the main families of p.o. in the region of the
radial and vertical ILRs in section~\ref{sec:between}. In
section~\ref{sec:x1v1p} we investigate the role of chaotic orbits in the side-on
profiles based on the dynamics of the x1v1 family, while the different
mechanisms that could lead to an `\emp'' or 
\begin{large}``$\infty$''\end{large}-type profile are considered in 
section~\ref{sec:x1v2}. In section~\ref{sec:geom} we deal with projection
effects that are important for the profiles we examine and finally we
give a summary of the paper by enumerating our conclusions in
section~\ref{sec:concl}.

\section{The model}
\label{sec:model}
Since clear b/p bulges and especially {\sf X} features are usually associated
with strong bars we have chosen the ``strong bar case'' model of \cite{spa02b}
for our calculations. This model has the most massive bar with respect to the
disc from all other models investigated in that paper. The general model
consists of a Miyamoto disc, a Plummer bulge and a 3D Ferrers bar. It is a
popular and extensively studied model for 3D bars, initially used by
\cite{pf84}. Here, we briefly describe the three components of the model to
facilitate the reader of the current paper.

The potential of a Miyamoto disc (Miyamoto \& Nagai 1975) reads:
\begin{equation}
\label{potd}
\Phi _{D}=-\frac{GM_{D}}{\sqrt{x^{2}+y^{2}+(A+\sqrt{B^{2}+z^{2}})^{2}}},
\end{equation}
where \( M_{D} \) is the total mass of the disc, $A$ and $B$ are the
horizontal and vertical scale lengths, and $G$ is the gravitational
constant. 

For the bulge we used a Plummer sphere \citep{pl11} with potential:
\begin{equation}
\label{pots}
\Phi _{S}=-\frac{GM_{S}}{\sqrt{x^{2}+y^{2}+z^{2}+\epsilon _{s}^{2}}},
\end{equation}
where \( \epsilon _{s} \) is the scale length of the bulge and \(
M_{S} \) is its total mass. 

Finally, the density \( \rho (x) \) of a triaxial Ferrers bar is:
\begin{equation}
\label{densd}
\rho =\left\{ \begin{array}{lcc}
\displaystyle{\frac{105M_{B}}{32\pi abc}(1-m^{2})^{2}} & {\mbox for} &
m \lid 1\\ 
 & & \\
\displaystyle{0} & {\mbox for}  & m>1
\end{array}\right. ,
\end{equation}
where
\begin{equation}
\label{semiaxis}
m^{2}=\frac{y^{2}}{a^{2}}+\frac{x^{2}}{b^{2}}+\frac{z^{2}}{c^{2}}\, \, ,\, \,
 \, 
 a>b>c,
\end{equation}
\( a \), \( b \), \( c \) are the semi-axes and \( M_{B} \) is the
mass of the bar component. The corresponding potential \( \Phi _{B} \)
and the forces are given in Pfenniger (1984). 

For the Miyamoto disc we use A=3 and B=1, and for the axes of the Ferrers bar we
set $a:b:c = 6:1.5:0.6$, as in \cite{pf84} and \cite{spa02b}. The masses of the
three components satisfy \( G(M_{D}+M_{S}+M_{B})=1 \). The length unit is taken
as 1~kpc, the time unit as 1~Myr and the mass unit as $ 2\times 10^{11}
M_{\odot}$. If not otherwise indicated, the lengths mentioned all over the text
and in the figures are in kpc. Units will be not repeated on the axes of the
figures. 

The total potential of our model is given by 
\begin{equation}
\Phi(x,y,z)= \Phi_D + \Phi_S + \Phi_B
\end{equation}

The 3D bar is rotating around its short $z$ axis. The $x$ axis is the
intermediate and the $y$ axis the long one. The system is rotating with an
angular speed $\Omega_{b}$ and the Hamiltonian governing the motion of a
test-particle can be written in the form:

\begin{equation}
H= \frac{1}{2}(p_{x}^{2} + p_{y}^{2} + p_{z}^{2}) +
    \Phi(x,y,z) - \Omega_{b}(x p_{y} - y p_{x})    ,
\end{equation}
where $p_{x},~ p_{y},$ and $p_{z}$ are the canonically conjugate momenta. We
will hereafter denote the numerical value of the Hamiltonian by E$_J$ and
refer to it as the Jacobi constant or, more loosely, as the `energy'.
The corresponding equations of motion are:
\begin{eqnarray}
\dot{x}=p_{x}+\Omega_{b}y, & \dot{y}=p_{y}-\Omega_{b}x, &
\dot{z}=p_{z} \nonumber \\
\dot{p_{x}}= -\frac{\partial \Phi}{\partial x} + \Omega_{b}p_{y}, &
\displaystyle{\dot{p_{y}}=-\frac{\partial \Phi}{\partial y} - \Omega_{b}p_{x},}
& 
\dot{p_{z}}=-\frac{\partial \Phi}{\partial z}
\end{eqnarray}

In Table~\ref{tab:models} we summarise the parameters of the specific model we
will use in our paper.
 
\begin{table*}
\caption[]{The parameters used in our model. G is the gravitational constant,
M$_D$, M$_B$, M$_S$ are the masses of the disk, the bar and the bulge
respectively, $\epsilon_s$ is the scale length of the bulge, \O$_{b}$ is the
pattern speed of the bar, E$_J$(r-ILR) and E$_J$(v-ILR) are the values of the
Jacobi constant for the radial and vertical 2:1 resonances, given with higher
accuracy than in \cite{spa02b}, and $R_c$ is the corotation radius.}
\label{tab:models}
\begin{center}
\begin{tabular}{ccccccccc}
  GM$_D$ & GM$_B$ & GM$_S$ & $\epsilon_s$ & \O$_{b}$ & E$_J$(r-ILR)
 & E$_J$(v-ILR) & $R_c$ \\ 
\hline
  0.72 &  0.2  & 0.08 & 0.4 &  0.054 & -0.467878 & -0.438317& 6.31 \\

\hline
\end{tabular}
\end{center}
\end{table*}

\subsection{Periodic orbits and their stability}
\label{sec:po}
Most information about the dynamics of our Hamiltonian system is given by the
study of the main families of p.o. that belong to the x1 tree \citep{spa02a}.
The phase space we study is structured by the p.o. and the way it is
structured depends on their stability. 

The space of section in the case of a 3D system is 4D. In order to find a p.o.,
the equations of motion are solved for a given value of the Hamiltonian,
starting with initial conditions $(x_{0},p_{x_0},z_{0},p_{z_0})$ in the plane
$y$=0, for $p_{y} > 0$. The exact initial conditions for the periodic orbit are
calculated using a Newton iterative method. A periodic orbit is found when the
initial and final coordinates coincide with an accuracy at least 10$^{-11}$. 
For the integration of the orbits we used a fourth order
Runge-Kutta scheme. The relative error in the energy remains in our
calculations less than 10$^{-15}$.

For the estimation of the linear stability of a periodic orbit we first consider
small deviations from its initial conditions, and then integrate the perturbed
orbit to the next upward intersection. In this way a transformation $T:
\mathbf{R}^{4} \to \mathbf{R}^{4}$ is established, which relates the initial
with the final point. The relation of the final deviations of this neighbouring
orbit from the periodic one, with the initially introduced deviations can be
written in vector form as: $\vec{\xi}=M\,\vec{\xi_{0}}$. Here $\vec{\xi}$ is the
final deviation, $\vec{\xi_{0}}$ is the initial deviation and $M$ is a $4 \times
4$ matrix, called the monodromy matrix.  It can be shown that the characteristic
equation is written in the form $\lambda^{4} + \alpha \lambda^{3} + \beta
\lambda^{2} + \alpha \lambda + 1 = 0$. Its solutions $(\lambda_i, i=1,2,3,4)$
obey the relations $\lambda_{1}\,\lambda_{2}=1$ and $\lambda_{3}\,\lambda_{4}=1$
and for each pair we can write:
\begin{equation}
\lambda_{i}, 1/\lambda_{i} = \frac{1}{2} 
[-b_{i}\pm(b_{i}^{2} -4)^{\frac{1}{2}}],
\label{eq:eigen}
\end{equation}
where $\displaystyle b_{i} = 1/2\,( \alpha \pm \Delta^{1/2})$ and 
\begin{equation}
\Delta =\alpha^{2} - 4 (\beta - 2).
\label{eq:delta}
\end{equation}

The quantities $b_{1}$ and $b_{2}$ are called the stability indices. One of them
is associated with radial and the other one with vertical perturbations. If
$\Delta > 0$, $|b_{1}|<2$ and $|b_{2}|<2$, the 4 eigenvalues are on the unit
circle and the periodic orbit is called `stable'. If $\Delta > 0$, and
$|b_{1}|>2$, $|b_{2}|<2$, or $|b_{2}|>2$, $|b_{1}|<2$, two eigenvalues are on
the real axis and two on the unit circle, and the periodic orbit is called
`simple unstable'.  If $\Delta > 0$, $|b_{1}|>2$, and $|b_{2}|>2$, all four
eigenvalues are on the real axis, and the periodic orbit is called `double
unstable'.  Finally, $\Delta < 0$ means that all four eigenvalues are complex
numbers but {\em off} the unit circle. In this case the orbit is characterised
as `complex unstable' \citep{cm85, dh85, pf85a, pf85b, z93}. We use the symbols
S, U, DU, $\Delta$ to refer to $stable$, $simple~unstable$,
$double~unstable$ and $complex~unstable$ periodic orbits respectively. 

The distinction of the various types of instability has been initially presented
by \cite{br69} and \cite{h75}, and has been used in studies of the stability of
periodic orbits in systems of three degrees of freedom. For an extended
description the reader may refer to \cite{pf84} and \cite{cm85}. For a general
discussion of the kinds of instability encountered in Hamiltonian systems of
{\sf N} degrees of freedom the reader may refer to \cite{sk01}.

The `stability diagram' is a diagram that describes the stability of a family of
periodic orbits in a given potential when one of the parameters of the system
varies (e.g. the numerical value of the Hamiltonian E$_J$), while all other
parameters remain constant \citep{cb85, cm85}. In other words it gives the
variation of the stability indices $b_1$ and $b_2$ as the energy, or any other
parameter, varies. The resulting curves are called the ``stability curves''.
Such a diagram helps in finding the transitions from stability to instability or
from one to another kind of instability.  Intersections, or tangencies, of a
stability index with the $b=-2$ axis are of special importance for the dynamics
of the system. When this happens a new family is generated by bifurcation from
the initial one and has the same multiplicity with it. The new family of p.o.
inherits also the kind of stability that characterises the parent family.
Intersections, or tangencies, of the stability curves with the $b=2$ axis, also
generate new families but are accompanied by period doubling. The central family
of the system is the well known x1 family \cite[see e.g][]{cg89}, which is
reduced to circular orbits on the equatorial plane in the axisymmetric part of
the potential. In the full potential the members of this family are p.o. of
elliptical shape, elongated along the major axis of the bar, remaining always on
the equatorial plane. The ``x1-tree'' is composed by the x1 family itself
together with the 3D families, which are generated at specific S$\rightarrow$U
or U$\rightarrow$S transitions. At these transitions the stability index $b_2$
associated with the vertical perturbations has two intersections, or a tangency,
with the $b=-2$ axis. They happen at the regions of the vertical resonances
\citep[see e.g][]{pg96}. The ``x1-tree'' is the backbone of b/p bulges and {\sf
X} features \citep{psa02}. In standard models the first S$\rightarrow$U and
U$\rightarrow$S transitions due to an intersection of $b_2$ with the $b=-2$ axis
are encountered near the vertical 2:1 resonance and the bifurcated simple
periodic 3D families of p.o. of the x1-tree are introduced in pairs. Following
the nomenclature of the fiducial case in \cite{spa02a} these families are called
x1v1 and x1v2 (associated with the vertical 2:1 resonance), x1v3 and x1v4
(associated with the vertical 3:1 resonance) etc. Finally we have to remark that
at the transition to complex instability we have in general no bifurcating
families of \textit{periodic} orbits. 

\begin{figure*}
\begin{center}
\resizebox{180mm}{!}{\includegraphics[angle=0]{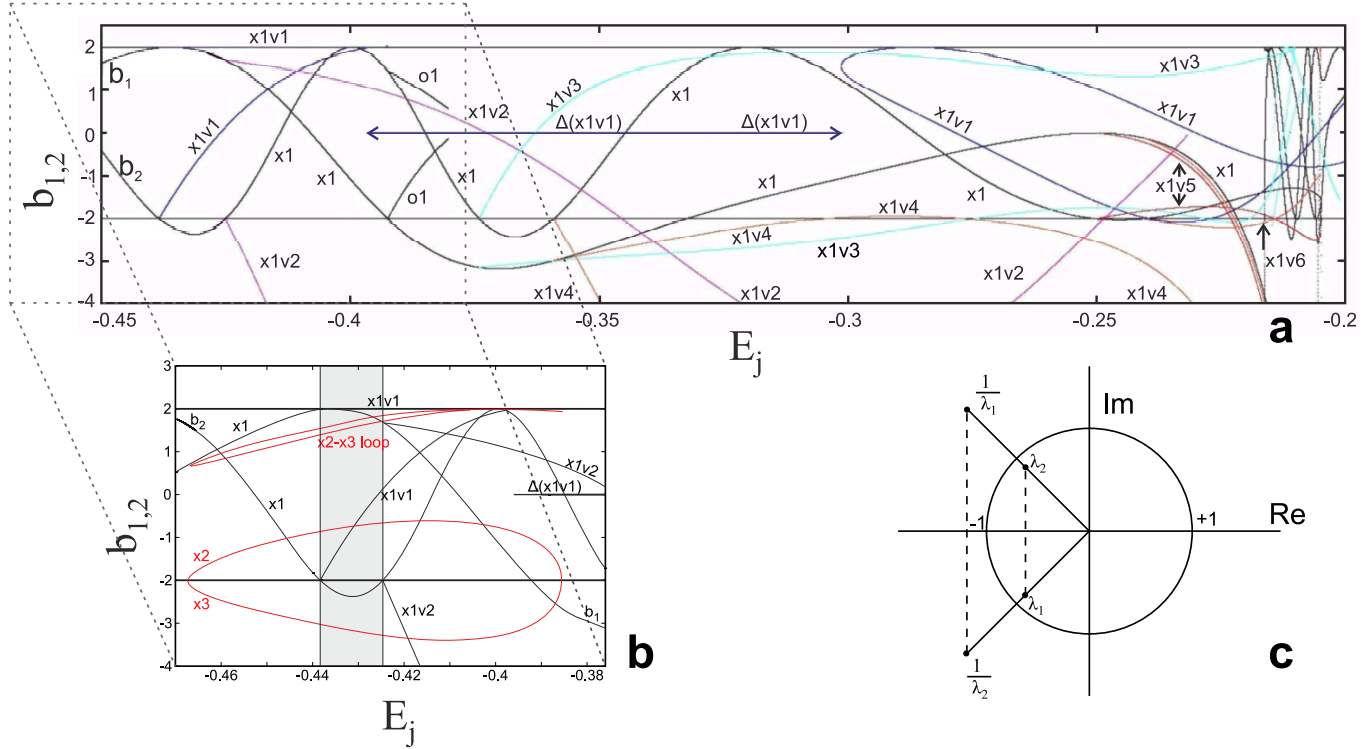}}
\end{center}
\caption{(a) The evolution of stability for x1 and the families of the x1 tree
as the energy varies. The curves existing already for \ej$ = -0.45$ belong to
the x1 family. Index $b_1$ is associated with radial, while $b_2$ with vertical
perturbations. $\Delta$(x1v1) marks the energy interval in which x1v1 is complex
unstable. (b) Stability diagram for the main families of p.o. at the radial and
vertical 2:1 resonance region. The indices associated with the x2-x3 loop are
plotted with red. In the grey shaded region $\Delta$E$_{vILR}$ (see text) x1 is
vertically unstable. (c) The four eigenvalues of the family x1v1 $(\lambda_{i},
1/\lambda_{i}, i=1,2)$ in its complex unstable part.}
\label{stabdiag} 
\end{figure*}

The stability diagram of our model that describes the evolution of stability for
all families of the x1-tree as \ej varies is given in Fig.~\ref{stabdiag}a. The
two stability curves, that exist already for \ej$ \leqq -0.45$ (left side of the
figure), belong to x1. The stability index $b_1$ is associated with the radial,
while $b_2$ with the vertical perturbation. The families x1v1, x1v2 etc. are
introduced at the intersections of $b_2$ with the $b=-2$ axis. In
Fig.~\ref{stabdiag}b we zoom into the ILR region, which is of special importance
for the b/p dynamics. We will refer to it in detail in
section~\ref{sec:between}. The straight line segments in Fig.~\ref{stabdiag}a
and b at $b_{1,2}=0$ for \ej$\gtrapprox -0.3965$, labelled ``$\Delta$(x1v1)'' is
drawn to mark the area over which the x1v1 family is complex unstable. In this
region the inidices $b_1$ and $b_2$ in Eq.~\ref{eq:eigen} cannot be defined
since we have $\Delta < 0$ in Eq.~\ref{eq:delta}. The location of the four
eigenvalues of the x1v1 p.o. with respect to the unit circle is as shown in
Fig.~\ref{stabdiag}c. We will refer to Fig.~\ref{stabdiag} throughout the
text, describing the dynamical behaviour in the neighbourhood of p.o.
participating in the building of the b/p bulge. 

\section{Where does the peanut start?}
\label{sec:start}
The radial ILR region of the
model is found from the energy interval over which the families x2 and x3
\citep{cg89} exist. E$_J$(r-ILR) in Table~\ref{tab:models} is taken as the left
limit of the x2-x3 loop in the characteristic or in the stability diagram.
E$_J$(v-ILR) is taken at the energy where we have the S$\rightarrow$U transition
of x1 and the introduction in the system of the x1v1 family. The first
(vertical) instability strip in the energies of x1 occurs at $-0.438317 < $\ej
$< -0.424945$ (Fig.~\ref{stabdiag}a). One more 3D family (x1v2) associated with
the vertical ILR starts at the right border of this interval where we have the
U$\rightarrow$S transition (Fig.~\ref{stabdiag}a). 

It is generally believed that the peanut starts at the vertical 2:1 (v-ILR)
resonance, where we encounter 3D p.o. with elliptical projections on the
equatorial plane and ``smiles'' or ``frowns'' in their side-on views, i.e. p.o.
of the x1v1 family \citep{pf84, cetal90, psa02, betal06, px06, m-vetal06}. We
note that \citet{cetal90} emphasise that at the end of their $N$-body models the
radial and vertical 2:1 resonances coincide.  

The bar is elongated along the y-axis and its backbone is the x1-tree. 
The members of the x1v1 family are not the first 3D orbits in the system in
general. There
are other 3D families of p.o. in the very central part of the model associated
with the bulge component of the potential, as well as 3D non-periodic orbits
that support the disc in the same region. Among them there are 3D quasi-periodic
orbits with \ej$ < -0.438317$ associated with the planar x1 family. We
investigate here their contribution to the 3D shape of the bar in the central
region of the model by examining their effect in the resulting side-on profiles.

\begin{figure*}
\begin{center}
\resizebox{160mm}{!}{\includegraphics[angle=0]{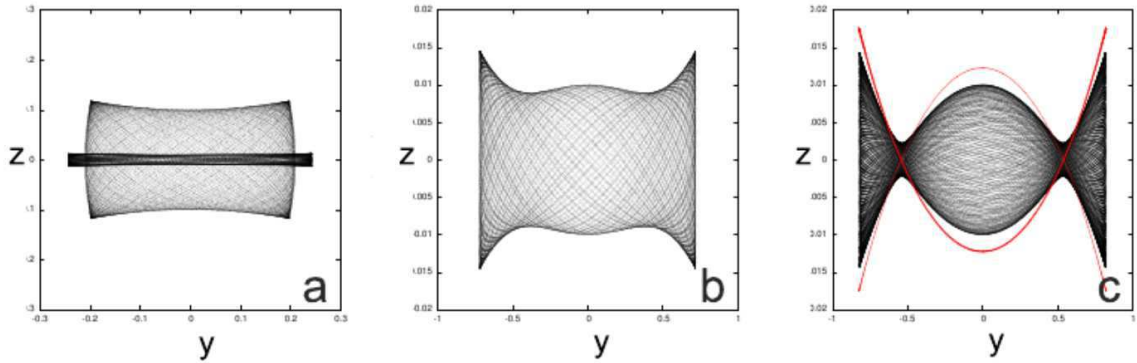}}
\end{center}
\caption{The (y,z) projections (edge-on) of quasi periodic orbits trapped around
x1 \textit{before} its first S$\rightarrow$U transition at the vertical 2:1
resonance. (a) for \ej$=-0.527983$, (b) \ej$=-0.45$, (c) \ej$=-0.438666$. The
two red curves in (c) are the projections of the p.o. x1v1 and its z-symmetric
p.o. at \ej$=-0.438225$, $after$ the S$\rightarrow$U transition of x1. Note the
different scales on the axes.}
\label{thickx1} 
\end{figure*}  
For \ej $ < -0.438317$, the only existing bar-supporting family of p.o.
belonging to the x1-tree is x1, which in this energy range is stable everywhere.
Albeit planar, x1 contributes to the building of the bar away from the
equatorial plane. Due to the stability of x1, quasi-periodic orbits around them
will be trapped in ``rotational'' or ``tube'' tori \citep{vetal97,kp11}, which
are 3D objects embedded in the 6D space. Perturbations of the initial conditions
of the x1 p.o. in the $z$ or $p_z$ direction will result to 3D orbits extending
in the 6D phase space.  

Interesting is the shape of the associated with these tori orbits in the
configuration space. We followed their morphological evolution as the
perturbations of the initial conditions of the p.o. at a certain energy increase
and also as the energy increases approaching the critical value, where we have
the S$\rightarrow$U transition of x1 at the vertical 2:1 resonance (\ej $
\approx -0.438317$).

In Fig.~\ref{thickx1}a we depict the $(y,z)$ projections of two quasi periodic
orbits around x1 at \ej$=-0.527983$. Both of them have the same $x_0$ initial
condition as the x1 p.o. at this energy $(x_0=0.072730404)$. The narrow, darker
one, is perturbed in the z direction by $\Delta z = 0.01$, while the thicker one
by $\Delta z = 0.1$. The projections of these two orbits have similar shapes. In
Fig.~\ref{thickx1}a the thick orbit has a maximum thickness about 0.2. The
energy of the two orbits is the same, thus the orbit with the higher $z$ has a
shorter extent along the $y$ direction. In the 3D configuration space these
orbits have a toroidal morphology. In the (y,z) projections they are
characterised by a minimum $z$ for $y=0$ and a maximum z at maximum $|y|$ as in
Fig.~\ref{thickx1}a. This is a typical, general morphology of the quasi-periodic
orbits around x1 away from the ILRs of the system.  

However, as energy increases we observe that initially for small perturbations,
the minimum $z$ for $y=0$ becomes a local maximum, as in the case of the (y,z)
projection of the orbit in Fig.~\ref{thickx1}b. In this particular case we have
an orbit with \ej $=-0.45$ having initial conditions that deviate from those of
the x1 p.o by $\Delta z$=0.01. In order to clearly observe the formation of the
local maximum, the scale of the z-axis is 0.02 times that of the y-axis.

By studying the quasi-periodic orbits around x1 for energies \textit{just
before} the S$\rightarrow$U transition of x1 at the vertical 2:1 resonance, we
realise that the local $z$ maximum at $y=0$ becomes prominent for quasi-periodic
orbits in the neighbourhood of x1. The orbit in Fig.~\ref{thickx1}c is for \ej
=$-0.438666$ and occurs by perturbing the x1 p.o. with initial conditions $(x_0,
z_0, p_x,p_z)$=(0.0930694109,0,0,0) by $\Delta z$=0.01. The projection of the
orbit has taken a shape like an ``\emp'' symbol. The p.o. x1 at this energy is
still stable ($b_1\approx -1.98$, $b_2\approx 1.98$). Despite the fact that we
are \textit{before} the S$\rightarrow$U transition at which the x1v1 family is
bifurcated, the shape of this orbit is typical of an orbital structure expected
to appear in the neighbourhood of the stable x1v1 family and its symmetric with
respect to the equatorial plane (x1v1$^\prime$) \citep{psa02}. There is a
connection between the evolution of the border of the quasi-periodic orbits and
the shape of the orbits of the family of p.o. \textit{to be} bifurcated. In
order to show this   we plot in Fig.~\ref{thickx1}c with red the projections of
the x1v1 and x1v1$^\prime$ p.o. for a slightly larger energy (\ej$=-0.438225$)
when they have been introduced in the system as stable p.o. $after$ the
S$\rightarrow$U transition of x1. 

\begin{figure}
\begin{center}
\resizebox{80mm}{!}{\includegraphics[angle=0]{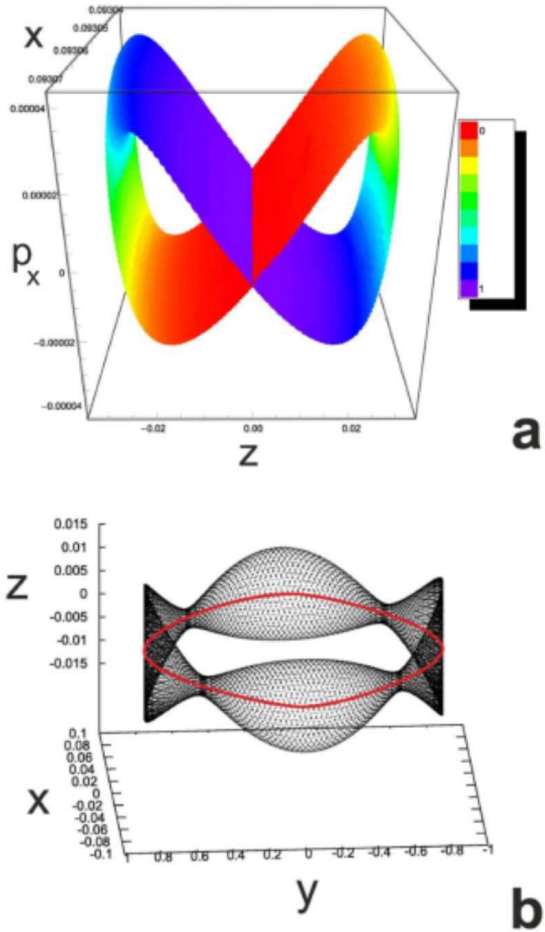}}
\end{center}
\caption{(a) A 4D representation of the Poincar\'{e} surface of section for the
quasi-periodic orbit with initial conditions $(x,z,p_x,p_z)$
= (0.0930694109, 0, 0, 0.01) at \ej$=-0.438666$. The spatial coordinates are
$(x,z,p_x)$, while the colour of the consequents represents their $p_z$
value. (b) A 3D view of the quasi-periodic orbit (black) in the configuration
space. The red curve is the x1 p.o. on the $z=0$ plane at the same energy.}
\label{cr_xzdz_43866} 
\end{figure}

On account of the stability of the x1 family, 3D quasi-periodic orbits are
expected to be trapped around their p.o. in invariant tori
called ``tube'' or ``rotational''.
\citep{vetal97, kp11}. By means of the colour-rotation method 
\citep{pz94}, these tori appear
in the case of family x1, before the v-ILR, as typical ``tube tori''. This
happens because there is a 3D projection, $(x,z,p_x)$, in which the torus
appears as folded and self-intersecting \citep{vetal97}. 
In Fig.~\ref{cr_xzdz_43866}a we depict the perturbed x1 orbit by $\Delta p_z =
0.01$ at \ej$=-0.438666$. For constructing this 4D object the orbit is
integrated for 500 periods of the x1 p.o. at the same energy. We use the
$(x,z,p_x)$ projection and the consequents are coloured according to their $p_z$
values, normalised in this case in the interval [0,1]. The colour bar, at the
right side of the figure, gives the colour variation. The point of view in
spherical coordinates is defined by the angles $(\theta,
\phi)=(270^{\circ},0^{\circ})$ and the distance of the observer from the centre
of the figure equals the largest dimension of the box surrounding the drawn
orbit (for more details see \cite{kp11}). The shape of the object resembles an
``8'' or ``$\infty$'' symbol. 
At the region of the intersection of the two
meeting branches we encounter colour values at the two opposite edges of the
spectrum, which means that it is a pseudo-intersection. 
The figure gives a
typical 4D Poincar\'{e} surface of section for all
3D bar-supporting quasi-periodic orbits before the first S$\rightarrow$U
transition of x1. A 3D view of this orbit in the configuration space is given in
Fig.~\ref{cr_xzdz_43866}b. The point of view in this figure is
$(\theta,\phi)=(60^{\circ},266^{\circ})$. The quasi-periodic orbit is drawn
with black. The red curve is the x1 p.o. on the $z=0$ plane at the same energy.
One can easily understand that by viewing this orbit side-on, the shape will be
as the one depicted in Fig.~\ref{thickx1}c. 
\begin{figure}
\begin{center}
\resizebox{80mm}{!}{\includegraphics[angle=0]{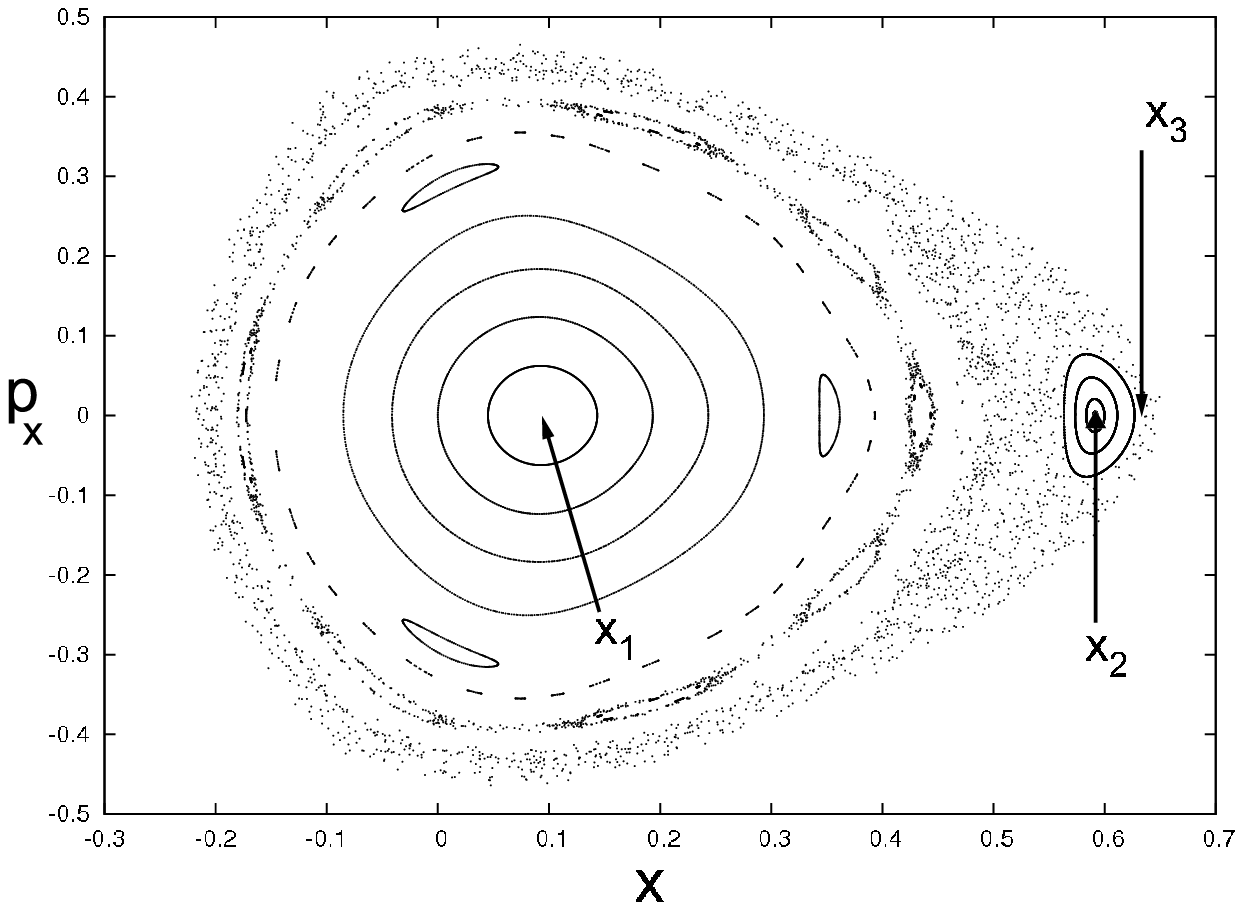}}
\end{center}
\caption{The $(x,p_x)$ cross section for $y=0, p_y>0$ for orbits on the $z=0$
plane for \ej$=-0.438225$. The location of the simple unstable p.o. x1 and x3,
as well as the stable p.o. x2 are indicated with arrows.}
\label{xxdot} 
\end{figure}  

The study of successive $(y,z)$ projections shows that the shape of the
quasi-periodic orbits with the same deviation from the initial conditions of x1
changes monotonically as the energy varies. 
In all
examined models we found particles in orbits supporting a b/p structure at the
region of the ILRs and \textit{before} the introduction of the x1v1 family in
the system. Based on the appearance of the local maximum at $y=0$ in the
$(y,z)$ projections, we can say that we have peanut-supporting orbits already 
at a distance of about 0.6~kpc along the major axis of the bar.

\section{The region of the radial and vertical ILRs}
\label{sec:between}
\subsection{Non-periodic orbits in the neighbourhood of x1}
Thus far we explored the contribution of the quasi-periodic orbits trapped by x1
to the formation of a peanut structure before the introduction in the system of
the x1v1 family. The x1 family is introduced in the system at the 2:1
\textit{vertical} resonance. The S$\rightarrow$U transition of x1, at \ej$
\approx -0.438317$ marks the beginning of the vertical ILR region, while in our
model the radial ILR already exists at this energy. Beyond \ej$ \approx
-0.438317$ the x1 family becomes simple unstable, with $b_2<-2$, for $-0.438317
<$ \ej\!\! $< -0.424945$. Let us call this energy interval $\Delta$E$_{vILR}$.
At the end of the  $\Delta$E$_{vILR}$ interval we have the U$\rightarrow$S
transition of x1. Within $\Delta $E$_{vILR}$ the family x1 is simple unstable
(U), while x1v1 is stable (S). We find also the two known planar families of
p.o., x2 as stable and x3 as simple unstable.   The energy range over which the
x2, x3 families  extend is significantly larger than the $\Delta $E$_{vILR}$
region. We get an overview of the interconnections of the various families of
p.o. in the region of the radial and vertical ILR by zooming into the evolution
of the stability indices of all implicated families (Fig.~\ref{stabdiag}b). We
note that in the scale we use for presenting Fig.~\ref{stabdiag}b (let alone
Fig.~\ref{stabdiag}a) the $b_1$ index of x1v1 appears coinciding with the $b=2$
axis. In reality $b_1$ just remains always very close to the $b=2$ axis, having
two tangencies at \ej $\approx -0.4104$ and at \ej $\approx -0.3982$. At \ej
$\approx -0.396$ $b_1$ meets $b_2$ and the x1v1 family becomes complex unstable.
The stability indices associated with the x2-x3 loop are plotted with red. 
 
In $\Delta $E$_{vILR}$ x1 is unstable relative to vertical perturbations,
because $b_2 < -2$. However, for perturbations \textit{on} the equatorial plane
x1 remains stable, because $-2<b_1<2$. A $(x,p_x)$ cross section in
$\Delta$E$_{vILR}$ with orbits deviating from x1 by $\Delta x$ or $\Delta p_x$,
and thus remaining on the equatorial plane, will be as the one in the example
depicted in Fig.~\ref{xxdot} for \ej$=-0.438225$. We note that despite the fact
that for this \ej both x1 and x3 p.o. are characterised as simple unstable (U),
the phase space structure in their neighbourhood, as expected, has different
properties, because the two eigenvalues \textit{on} the unit circle in the
former case are associated with the radial perturbations, while in the latter
with the vertical ones (see section \ref{sec:x2x3} below). 
\begin{figure*}
\begin{center}
\resizebox{160mm}{!}{\includegraphics[angle=0]{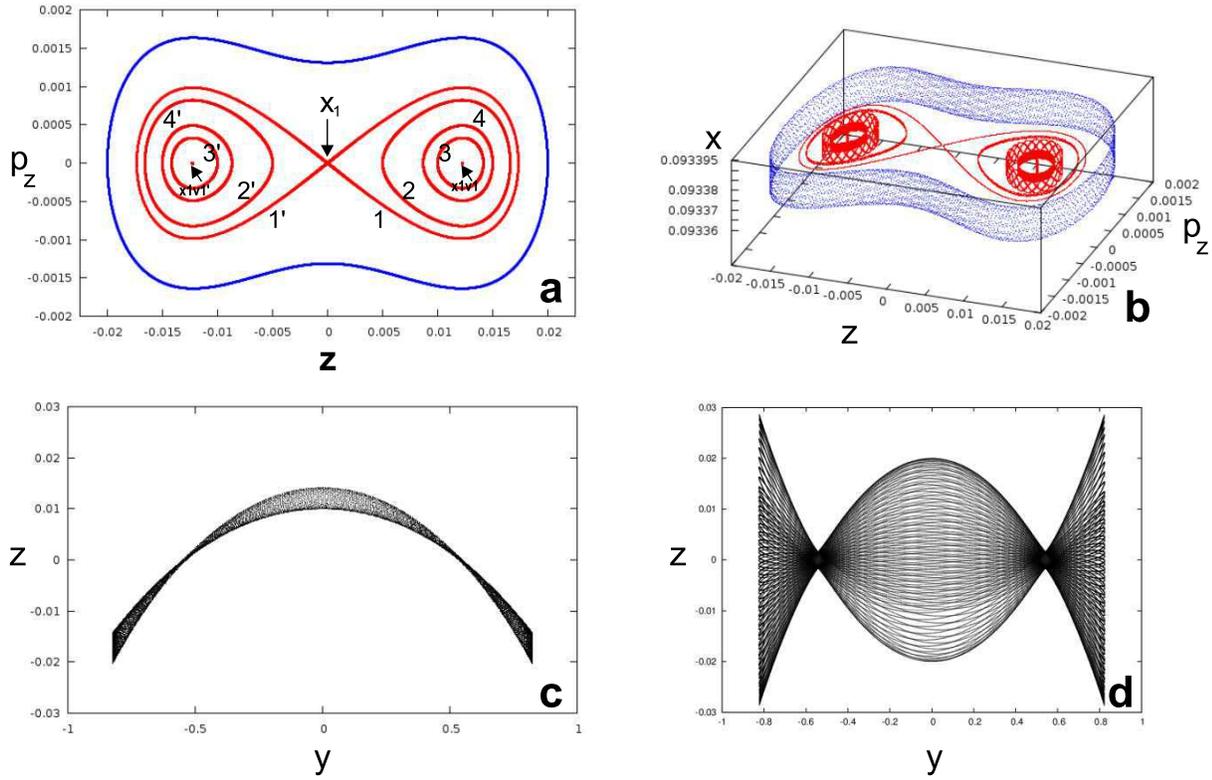}}
\end{center}
\caption{Perturbed x1 orbits at \ej=-0.438225. (a) The red curves numbered
``1'',``2'', ``3'' and ``4'', correspond to x1 orbits perturbed by  $\Delta z =
\pm 0.0001, \pm 0.005, \pm 0.01, \pm 0.015$. The blue curve is for the x1 orbit
perturbed by  $\Delta z = 0.02$. The locations of x1, x1v1 and x1v1$^{\prime}$
are indicated with arrows. The dots in the centres of the ``stability islands''
correspond to tiny tori around x1v1 and x1v1$^{\prime}$. (b) The same orbits as
in (a) in the 3D $(z,p_z,x)$ projection. (c) The $(y,z)$ projection of the
orbit numbered ``3'' in (a). (d) The $(y,z)$ projection of the
orbit on the torus surrounding all three p.o. in (a) and (b). It has a
peanut-like shape. }
\label{new} 
\end{figure*}  

As regards the dynamics in the neighbourhood of x1 away from the equatorial
plane, for energies close beyond the bifurcating point, as is \ej $=-0.438225$
(Fig.~\ref{new}), we find the following: The \textit{vertically} perturbed x1
orbits are characterised by \textit{slow} diffusion in phase space. The reason
is the presence of the invariant tori belonging to the stable family x1v1, which
has been already bifurcated from x1 at \ej$=-0.438317$. Despite the fact that
the presence of rotational or tube tori cannot restrict a region in the 4D space
of section, their proximity to the initial conditions of the U x1 p.o. hinders
considerably the diffusion of the consequents of perturbed x1 orbits in the
available phase space. There is a range of small deviations from the initial
conditions of x1, in $z$ and in $p_z$ directions, for which the integrated
orbits will be orbits on the invariant tori of the bifurcated stable x1v1 p.o.
at the same energy. This is due to the fact that close beyond the critical
energy of the bifurcation of x1v1 from x1, the bifurcated family has almost the
same $x_0$ as x1 and a small $z_0 \neq 0$. In Fig.~\ref{new}a we give the
$(z,p_z)$ projection of the consequents of the x1 orbits perturbed successively
by $\Delta z = \pm 0.0001, \pm 0.005, \pm 0.01, \pm 0.015$ (red curves) and
$0.02$ (blue curve), at \ej=-0.438225. The location of x1 is at $(z,p_z)=(0,0)$.
The orbits on the red curves, for $\Delta z = \pm 0.0001$ to $0.015$, surround
the initial conditions of the p.o. x1v1 (right) and x1v1$^{\prime}$ (left). The
location of these p.o. is indicated with arrows (the plotted dots correspond to
tiny tori around them). The orbit with $0.02$ surrounds in the $(z,p_z)$ plane
all three p.o. In the 3D $(z,p_z,x)$ projection we observe (Fig.~\ref{new}b)
that all these orbits are on tori, the thickness of which ($x$-dimension)
increases with increasing $\Delta z$. The side-on morphology of the orbits on
the red curves is as the one in Fig.~\ref{new}c ($z_0=0.01$) for the closed
curves around x1v1 and symmetric to them with respect to the equatorial plane
for the closed curves around x1v1$^{\prime}$. The tori which surround all three
of the p.o., like the one with $z_0=0.02$, have a clear peanut structure,
similar to the one we have seen in Fig.~\ref{thickx1}c, at an energy
\textit{before} the S$\rightarrow$U transition of x1. This is given in
Fig.~\ref{new}d.

If we include in the vertical (by $\Delta z$, $\Delta p_z$) perturbations of x1
also $\Delta x$ or $\Delta p_x$ displacements, i.e. if we essentially perturb
vertically quasi-periodic orbits on closed curves around x1 in Fig.~\ref{xxdot},
the situation does not change considerably. For the region inside the fourth
closed curve around x1 in Fig.~\ref{xxdot}, there is again a range of vertical
perturbations for which our orbits abut on x1v1 invariant tori and the orbits in
the configuration space have a morphology similar to the orbits in
Figs.~\ref{thickx1}b,c. For example, orbits perturbed by $\Delta z \leq 0.1$
starting in this region belong to x1v1 invariant tori. Differentiations from
this behaviour is observed when we start integrating orbits with $z$ or $p_z
\neq 0$ starting at and beyond the fourth closed curve around x1 in the x1
``stability island'' of Fig.~\ref{xxdot}. There, for the same range of $\Delta
z$ or $\Delta p_z$ perturbations, for which we reached x1v1 tori before, we
encounter orbits with complicated morphologies that are not similar among
themselves as the perturbation varies. This indicates that we reached the limits
of the phase-space region occupied by the x1v1 tori in the 4D space of section. 
 
\begin{figure}
\begin{center}
\resizebox{83mm}{!}{\includegraphics[angle=0]{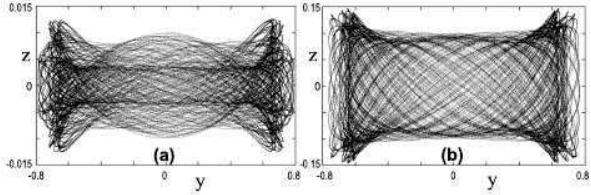}}
\end{center}
\caption{(y,z) projections of orbits starting on the 4th invariant curve around
x1 in Fig.~\ref{xxdot}, perturbed by $\Delta z = 0.01$ (a) and by $\Delta z =
0.1$ (b). Note the different scales of the axes.}
\label{othinv} 
\end{figure}

For instance, in Fig.~\ref{othinv} we show two such orbits with initial
conditions corresponding to perturbations of the orbit on the fourth invariant
curve around x1 in Fig.~\ref{xxdot}, perturbed by $\Delta z = 0.01$ (a) and by
$\Delta z = 0.1$ (b). The side-on profiles of these two orbits are not similar
among themselves. However, both of them could support a b/p morphology of a
central bulge. On one hand the side-on profile of the orbit in
Fig.~\ref{othinv}a reproduces a morphology resembling orbits trapped by either
branches of x1v1 (x1v1 and x1v1$^{\prime}$), which are symmetric with respect to
the equatorial plane.  However, it also differs from the orbit in
Fig.~\ref{new}d. On the other hand the orbit in Fig.~\ref{othinv}b has a x1
quasi-periodic orbit morphology similar to the one in Fig.~\ref{thickx1}a even
if we continue its integration for $10^5$ dynamical times. In \cite{pII} (paper
II), we will return to these orbits discussing orbits that combine boxy side-on
and face-on profiles.  For the time being we keep in mind that there is an
extended volume around x1 orbits, even in the energy interval where it is simple
unstable, that contributes to the boxiness of the side-on profile of the model. 

Beyond the fourth drawn elliptical curve around x1 in Fig.~\ref{xxdot}, we enter
a region where we have chains of islands of stability. This region on the
$(x,p_x)$ plane has been proven to be of special importance for the dynamics of
the b/p structures (see also paper II). Orbits with origin in this region may
have similar side-on morphologies with the orbit in Fig.~\ref{othinv}b for
thousands of dynamical times. Their sticky character is revealed very
illustrative by means of the colour-rotation method.  The 3D projections of the
space of section show us that these orbits remain confined for large times (of
the order of $10^2$ to $10^4$ dynamical times) in certain regions of the phase
space. We remind that usually in Galactic Dynamics we are interested about
time-scales of tens or in extreme cases a few hundreds of dynamical times. 

\begin{figure*}
\begin{center}
\resizebox{165mm}{!}{\includegraphics[angle=0]{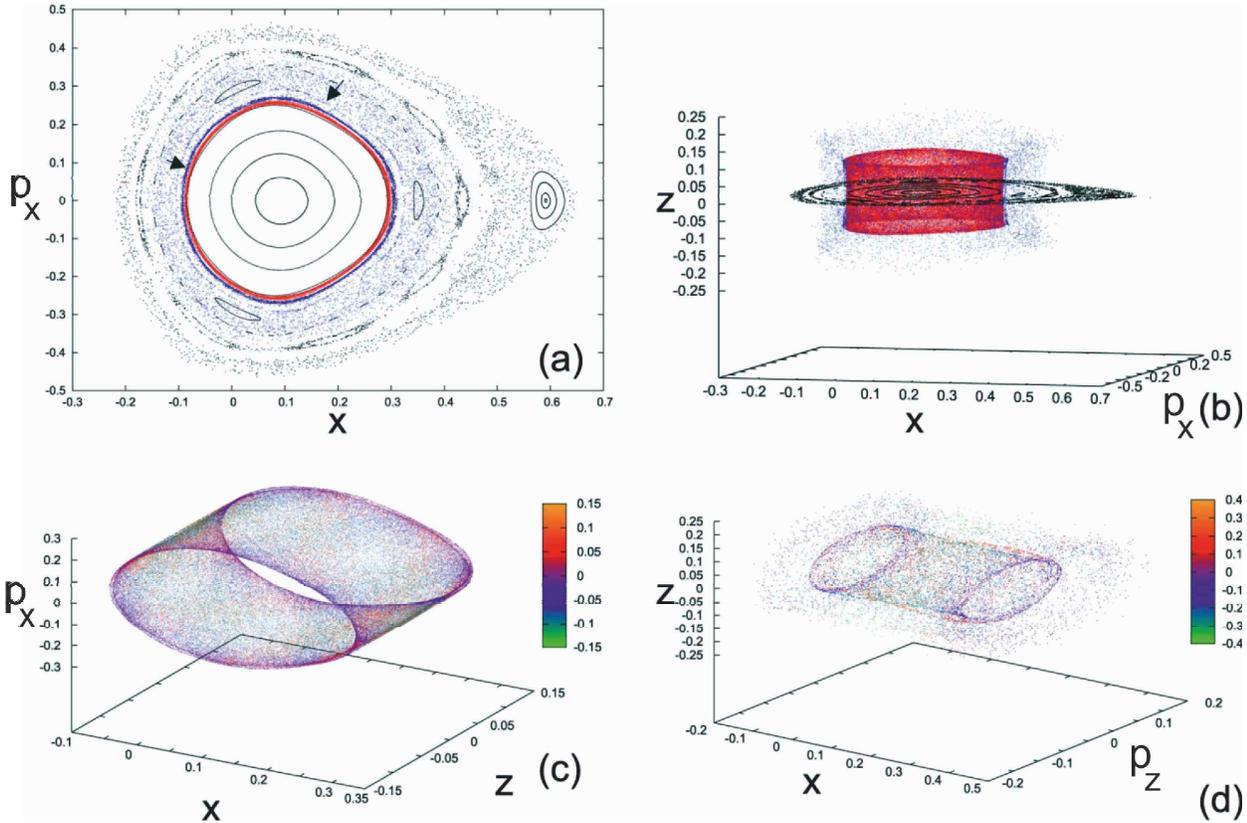}}
\end{center}
\caption{A chaotic orbit at \ej$=-0.438225$ trapped for thousands of dynamical
times of x1 in the region between the last drawn invariant curve around x1 and
a chain of eight stability islands in the $(x,p_x)$ projection of the 4D space
of section (Fig.~\ref{xxdot}). It is given with blue coloured consequents in (a)
and (b). In (c) and (d) we give 4D representations for the
``red'' and ``blue'' orbits of the panels (a) and (b) respectively.}
\label{brsticky} 
\end{figure*}

As an example, let us consider the orbit with initial conditions
$(x_0,z_0,p_{x_0},p_{z_0})=(0.304,0.1,0,0)$ and integrate it for $10^4$
dynamical times. In Fig.~\ref{brsticky}a we plot this orbit with blue points in
the $(x,p_x)$ projection on the surface of section presented in
Fig.~\ref{xxdot}. Fig.~\ref{brsticky}a  includes the invariant curves around x1
and x2 and the chaotic zone around both of them (black points). The points
coloured red  belong to the corresponding projection of the orbit in
Fig.~\ref{othinv}b. By careful inspection of Fig.~\ref{brsticky}a one can
observe that the blue consequents build at least two thin, relatively denser
layers with respect to their surrounding consequents in phase space. One of them
is found immediately after the ``red'' orbit and an even  thinner one further
out. We indicate these structures with black arrows. The orbit is trapped for
some time on these two layers before it expands and fills a larger region that
seems to be confined by a chain of islands just before we enter the chaotic
zone. By considering the $(x,p_x,z)$ projection (Fig.~\ref{brsticky}b) we
realise that the two orbits (``red'' and ``blue'') build in fact cylindrical
structures that remain confined also in the z-direction. This is more clear for
the consequents of the ``red'' orbit, which need an extremely large integration
time to diffuse in phase space (of the order of several Hubble times).
Nevertheless the chaotic character of both orbits is evident in their 4D
representations by means of the colour-rotation method. They are given in
Fig.~\ref{brsticky}c and Fig.~\ref{brsticky}d for the ``red'' and ``blue''
orbits respectively. In Fig.~\ref{brsticky}c we use $(x,z,p_x)$ as spatial
coordinates and we colour the consequents according to their $p_z$ values for
the ``red'' orbit. In Fig.~\ref{brsticky}d we use $(x,z,p_z)$ as spatial
coordinates and we colour the consequents according to their $p_x$ values for
the ``blue'' orbit. The colour bars giving the colours of the consequents are
given on the right side of these figures. Thus, the blue and red consequents are
not related to the colours we use in Figs.~\ref{brsticky}a, b for distinguishing
the two orbits. The first feature to be observed is that in both
Figs.~\ref{brsticky}c,d we have mixing of colours on the structures formed in
the 3D projections. All 4D combinations of spatial coordinates and colours give
qualitatively similar figures with those presented in Figs.~\ref{brsticky}c,d
for each one of the two orbits. In all 3D projections we have namely a toroidal
object with colour mixing. In Fig.~\ref{brsticky}d we observe that the ``blue''
orbit forms an inner toroidal structure before it expands, after about 5500
dynamical times, occupying a larger volume. The inner torus is tangent to the
one depicted in Fig.~\ref{brsticky}c as their $(x,p_x)$ projections in
Figs.~\ref{brsticky}a indicate. The part of the ``blue'' orbit corresponding to
the inner torus is morphologically very close to the one presented in
Fig.~\ref{othinv}b. 

In conclusion we can say that there is a plethora of sticky orbits in the
energy interval in which x1 is simple unstable, that can reinforce boxy side-on
profiles. In 3D projections of the 4D spaces of section, which include the
$(x,p_x)$ plane, these orbits are found trapped in cylindrical structures
between the closed curves around x1. This is an important dynamical mechanism
that has to be taken into account for the building of the peanut.

\subsection{The x2, x3 region}
\label{sec:x2x3}
\begin{figure}
\begin{center}
\resizebox{80mm}{!}{\includegraphics[angle=0]{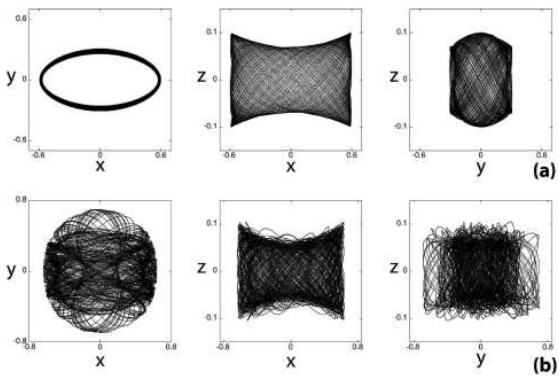}}
\end{center}
\caption{Orbits perturbed by $\Delta z=0.1$ from x2 (a) and from x3 (b)  for
\ej$=-0.438225$. They both contribute to boxy features in the central regions of
the bar. Orbits are integrated for 100 dynamical times.}
\label{x1x3per} 
\end{figure}

Apart from x1, the other simple unstable 2D family in the $\Delta $E$_{vILR}$
region is x3. Unlike x1 it is \textit{radially} unstable. On the $(x,p_x)$ plane
x3 is found inside a chaotic zone, which surrounds the closed curves around x1
and the stability island around x2 (Fig.~\ref{xxdot}). If we perturb by $\Delta
z$ the initial conditions of x3 we find in its neighbourhood orbits confined in
the z-direction of the configurations space. However, they are not as sharply
defined as the quasi-periodic orbit around x2. For comparison with x2, we give
the morphology of two perturbed by $\Delta z=0.1$ orbits, i.e. of the stable x2
and of the simple unstable x3 orbit within 100 dynamical times in
Fig.~\ref{x1x3per}a,b respectively. Note that in order to understand the
internal details of these orbits the scales of the axes in the (x,z) as well as
in the (y,z) projections are not equal. The perturbed x3 orbits spend more time
following a x2-flow (elongated perpendicularly to the bar) but during their
excursions in phase space away from the x2 island of stability their projections
on the equatorial plane precess. As a result they fill a disky region around a
box that extends along the x-axis (Fig.~\ref{x1x3per}b, left panel). At any rate
these orbits, like those in Fig.~\ref{othinv}, remain close to the equatorial
plane and support a boxy structure in the ILR region of the
model. 
Despite the difference in their stability character, the contribution of
non-periodic orbits around x3 to the local density away from the equatorial
plane is comparable with that of quasi-periodic orbits around x2. However, they
have a clearly different morphology in their face-on views.  

\begin{figure}
\begin{center}
\resizebox{82mm}{!}{\includegraphics[angle=0]{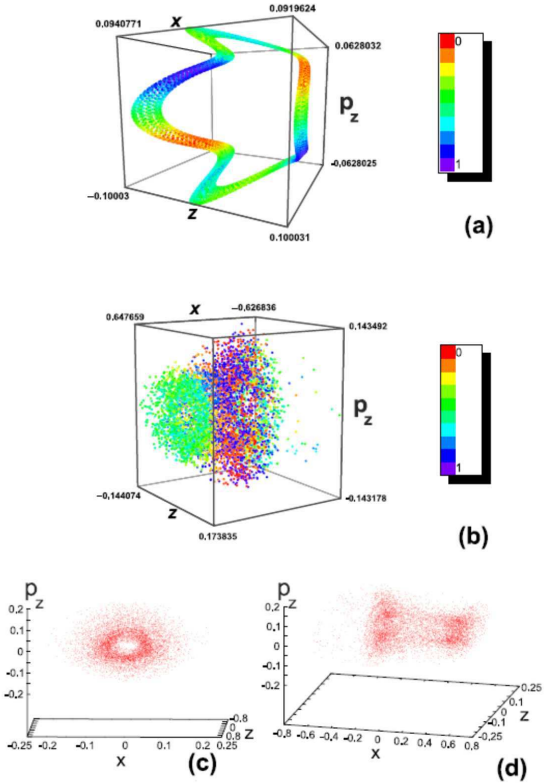}}
\end{center}
\caption{Comparison of the structures encountered in the phase space of two
simple unstable periodic orbits at \ej$=-0.438225$. (a) A 4D representation of
the surface of section in the neighbourhood of the vertically unstable x1 orbit
(b) a corresponding figure for the radially unstable x3. Panels (c) and (d)
clarify the geometry of the 3D projection of the orbit in (b). 
}
\label{sosx1x3} 
\end{figure}

Unlike in  the case of the x1 family, diffusion from the
immediate neighbourhood of x3 is not hindered by x1v1 tori. This is expected,
because x3 is not the parent family of x1v1 and so the tori of x1v1 are always
at a distance from the x3 p.o.
The different dynamical behaviours in the neighbourhood of the two 2D simple
unstable p.o. x1 and x3 is reflected in the different structures observed in the
4D surfaces of section of the perturbed orbits presented in Fig.~\ref{x1x3per}.
In order to get a clear view of these structures in the 4D cross section of the
phase space we have integrated them for a period 80 times larger than the one
used to depict them in the configuration space. Nevertheless, no new dynamical
features are introduced in the system during this much larger period. In
Fig.~\ref{sosx1x3}a we apply the colour-rotation method to visualise the 4D
space. In this particular case we use the $(x,z,p_z)$ projection and we colour
the consequents according to their $p_x$ variation, normalised in the [0,1]
interval. The x1 orbit is perturbed by $\Delta z=0.1$. We observe that its 4D
cross section is of the form of a ribbon on which colour has a smooth variation
according to the colour bar that is given at the right side of the figure. This
is a typical fingerprint of the 4D space of section in the neighbourhood of a
p.o. with this type of instability for times larger even than a Hubble time
\citep[cf with figure 3 in][]{kpc13}
On the other hand the corresponding 4D space of section of the perturbed x3
simple unstable orbit is characterised in its $(x,z,p_z)$ projection by two
conglomerations of consequents that are of toroidal form, as well as by a number
of consequents between them that form a bridge. This can be observed  in
Figs.~\ref{sosx1x3}b,c and d. In Fig.~\ref{sosx1x3}b we can see that the colour
on the front (left) toroidal structure has almost a single shade, reflecting the
proximity of the $p_x$ values in this region. The same is true for the toroidal
formation of the consequents at the back (right) side of the cross section. This
is less discernible in Fig.~\ref{sosx1x3}b due to the presence of the
consequents of the ``bridge'', which are projected on it and have mixed colours.
For this reason, in order to better understand the geometry of
Fig.~\ref{sosx1x3}b, we observe it from two more viewing angles in
Fig.~\ref{sosx1x3}c,d. This allows us to better understand the toroidal
character of the two dense regions (Fig.~\ref{sosx1x3}c) and the bridge between
them (Fig.~\ref{sosx1x3}d). In Fig.~\ref{sosx1x3}c, the left toroidal structure
of Fig.~\ref{sosx1x3}a is projected as a dense ring, being actually in front of
another, less dense, toroidal structure. We observe in the centre of the figure
an empty region, which corresponds to the hole of the two toroidal objects. In
Fig.~\ref{sosx1x3}d the dense regions (left and right) are the
``tori'' while the consequents between them belong to the bridge. We conclude
that the phase space in the neighbourhood of the two simple unstable 2D p.o. x1
and x3 is qualitatively very different. It is determined by the presence of
other families of p.o. co-existing in the same energy. In the case of x3 we
have a ``type1'' (tangent) bifurcation \citep[][section 2.4.3]{gcobook} with the
stable counterpart being the 2D family x2, while in the case of x1 we have a
``type 3'', direct bifurcation of the 3D families x1v1 and x1v1$^{\prime}$.
   
 
Despite the small size of the orbits investigated in this section, compared with
the dimensions of the b/p bulge, the orbits in $\Delta
$E$_{vILR}$ give an illustrative example of how the dynamical phenomenon of
stickiness becomes essential for building the peanut as \ej increases
approaching corotation. Similar phenomena take place at larger \ej\!\! as will
be described below. Stickiness is in many cases a significant dynamical
phenomenon in the neighbourhood of simple unstable p.o. A detailed study of
the sticky chaotic orbits just beyond the first $S\rightarrow U$ transition is
given by \cite{kp11} and by \cite{kpc13} in a
double Miyamoto-Nagai rotating potential. Here we find similar results and this
indicates a general dynamical behaviour in the phase space at the vertical ILR
region of rotating galactic potentials. 

\section{x1v1 peanuts}
\label{sec:x1v1p}
Up to now we examined the starting point and the central structure of the
peanut. However, our main goal is to study the overall structure of b/p bulges
and the {\sf X} feature. For this, most important are the outermost orbits that
determine to a large degree the observed morphology. As mentioned in the
introduction, the orbits that have been proposed by most authors of relevant
studies as building blocks belong to the stable family of p.o. that is
bifurcated at the vertical 2:1 resonance, i.e. the x1v1 family. We want to
investigate in what degree x1v1 is sufficient in building peanut and {\sf
X}-shaped bulges with dimensions similar to those observed in edge-on disc
galaxies. An obvious obstacle with this family is its complex
instability character over large energy intervals. This is a typical property of
this family in 3D rotating galactic potentials \citep{pf85b, pz90, spa02b,
kpc11}. 

In our model the longest projection of x1-tree p.o on the major axis of the bar
(y-axis) reaches a radius $y\approx 4.6$. This can be roughly considered
as the length of the bar that can be built by the orbits of our model. The bar
extends to 1.5 horizontal scale lengths of the Miyamoto disc. This is a value
close to the mean bar length as fraction of the exponential disc scale length
proposed by \citet{e05} for early type barred galaxies\footnote{This
ratio is usually larger in $N$-body simulations than the corresponding quantity
estimated for real galaxies \citep{e05} - see e.g. models that develop peanuts 
in \citet{am02} and in \citet{sn13} - and may change
during the life of a galaxy \citep{chetal13}}.

\begin{figure*}
\begin{center}
\resizebox{100mm}{!}{\includegraphics[angle=0]{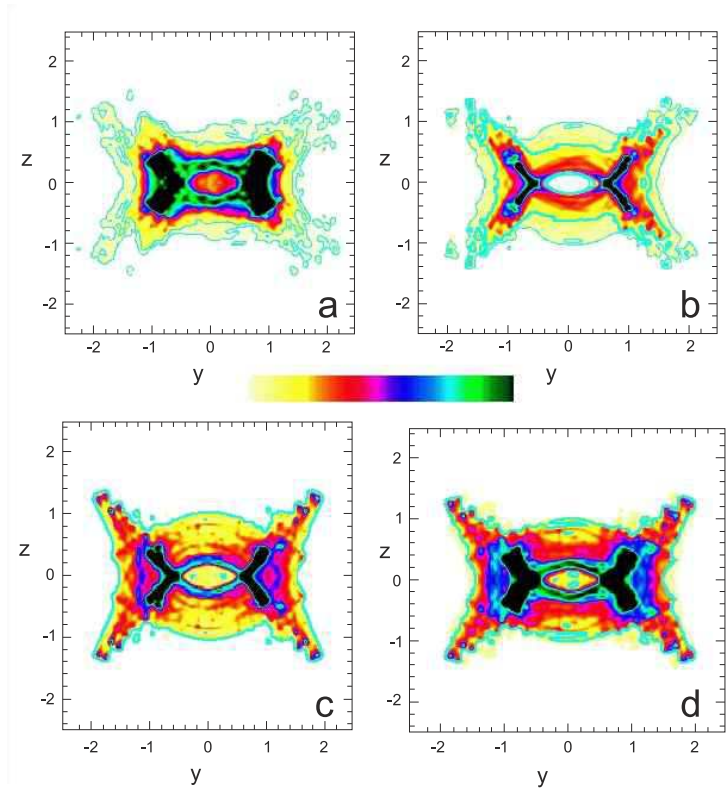}}
\end{center}
\caption{Side-on profiles built by perturbed x1v1 orbits. In (a) the imposed
perturbations are in the $z$ direction, in (b) in $x$, in (c) in $p_z$ by
0.05 and in (d) in $p_z$ by 0.1. Characteristic contours are overplotted...}
\label{prof} 
\end{figure*}

The maximum length of the b/p bulge in the side-on projection is expected
to be according to \cite{ldp00} $R_{max}\approx 1.7$ (see
Sect.~\ref{sec:intro}). Then, by measuring the dimensions of the {\sf X}
features in \cite{ar06} we estimate a height compatible with observations about
$z_{max} \approx 1.2$. This height is reached by x1v1 p.o. at \ej\!\!=$-0.27$.
Taking this into account we have built indicative profiles with non-periodic
orbits in the neighbourhood of the x1v1 p.o. Such side-on, $(y,z)$, profiles are
given in Fig.~\ref{prof}. They have been constructed by taking eight orbits at
equally spaced energy intervals between $-0.41 \leq $\ej$ \leq -0.27$ and
integrating them for time equal to 12 periods of the x1v1 p.o. at  the energy
\ej\!\!=$-0.27$. The profiles have been converted to images by means of the
ESO-Midas software. The colour bar in the middle of the figure indicates
increasing density from left to right. The criterion used for selecting these
orbits in each panel differs. In Fig.~\ref{prof}a we have taken perturbed x1v1
orbits by $\Delta z$. The perturbations were such as to secure a clearly by eye
discernible ``\emp'' feature formed when we plot the side-on views of
the non-periodic orbits and their symmetric with respect to the equatorial plane
of the model. This  $\Delta z$ perturbation is less than 0.1 pc and is added to
the $z_0$ initial condition of each p.o. x1v1. The rest of the initial
conditions of the used orbits are 0, as for the corresponding x1v1 p.o. The
orbital profile in Fig.~\ref{prof}b is constructed by considering perturbations
by $\Delta x$. This time we obtain ``\emp'' morphologies for $\Delta x$
perturbations of the individual orbits in the range $0.05 < \Delta x < 0.25$. In
Figs.~\ref{prof}c,d we start our orbits with $p_{z}$=0.05 and $p_{z}$=0.1
respectively, while the initial $p_{z_0}$ component for orbits of the x1v1
family is 0. In all cases the individual orbits build in their $(y,z)$ (side-on)
projections a ``\emp''-type profile. However, we have to stress that, like in
the case of the profiles with periodic orbits presented in \cite{psa02} and
\cite{px06}, the wings of the {\sf X} in the profiles in Fig.~\ref{prof} are not
along the wings of the individual orbits. In other words by considering all
orbits together we do not form a single ``\emp'' structure. The wings of the
{\sf X} features in the profiles are formed along the z-maxima of the orbits,
where $p_{z}=0$ (cf with figure 19a in \cite{psa02}). This will become apparent
also in the following section, Sect.~\ref{sect:compl}. For the time being we
conclude that x1v1 p.o. perturbed by $\Delta x$ or by $\Delta p_z$, with initial
$z_0 = 0$ form a sharper {\sf X} profile than the orbits starting away
from the equatorial plane $z_0 \neq 0$. We underline the fact that by
constructing these profiles we have applied the perturbations to the x1v1 p.o.
independently of their stability. 

\subsection{Contribution by orbits close to complex instability}
\label{sect:compl}
Qualitatively, the orbits that participate in the building of ``x1v1-based''
{\sf X} features reaching heights up to $z=1.2$ are of two kinds. They are
either quasi-periodic orbits trapped by stable x1v1 p.o. or chaotic orbits
associated with complex instability. The former are encountered at energies
$-0.438 <$ \ej $< -0.396$ and \ej $> -0.301$, while the latter at $-0.396 <$ \ej
$< -0.301$. The quasi-periodic orbits shape the innermost, as well as the
outermost, and thus most important for the overall morphology, part of the
peanut. On the other hand, over the larger part of the \ej interval, the x1v1
family is complex unstable. Nevertheless, our investigation shows that even
if we integrate the chaotic orbits participating in Fig.~\ref{prof} for times
equal to 40 periods of the outermost x1v1 p.o. participating in
the {\sf X} structure the boxy morphology is not destroyed.
 
\begin{figure*}
\begin{center}
\resizebox{140mm}{!}{\includegraphics[angle=0]{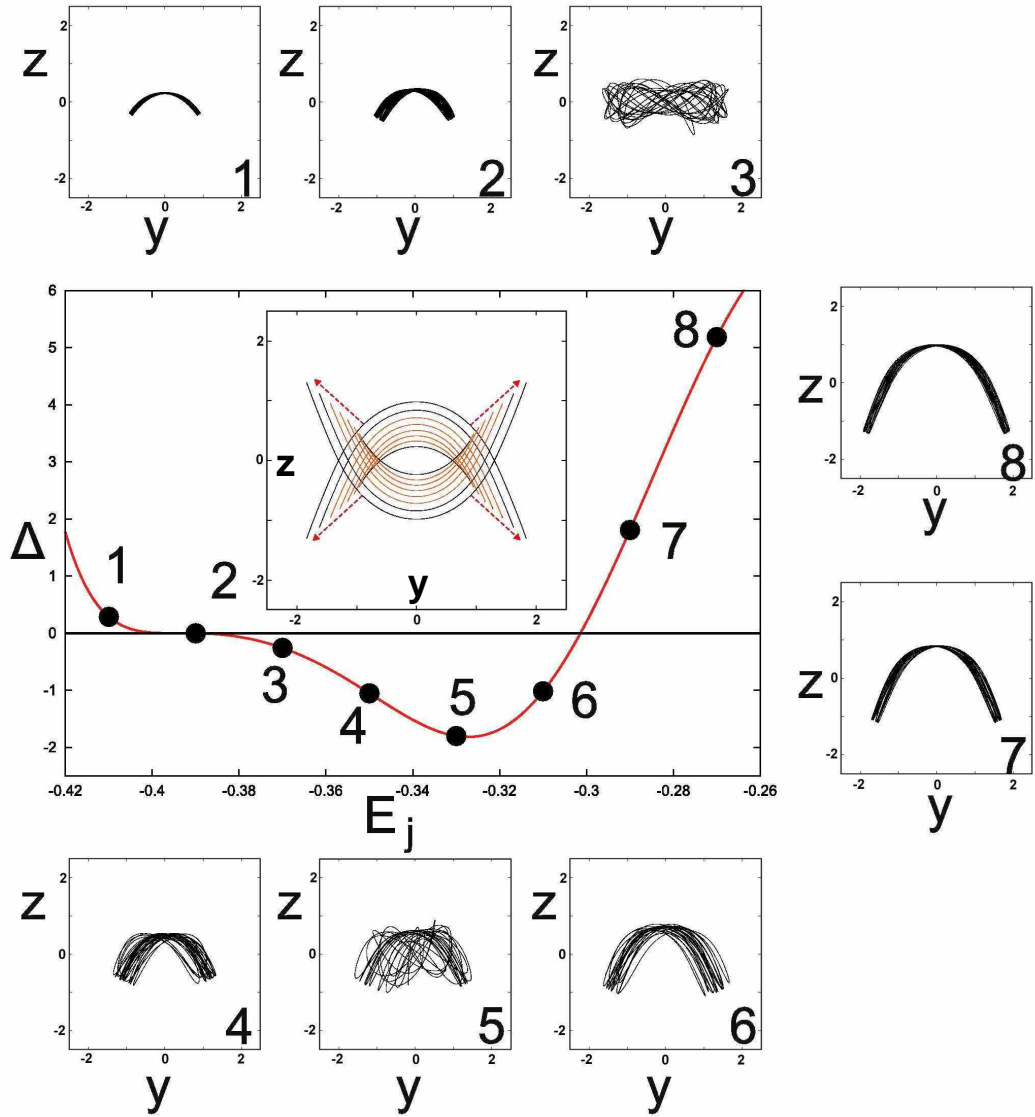}}
\end{center}
\caption{In the central frame we give the variation of $\Delta$ with the energy,
for 8 x1v1 p.o. The embedded frame depicts the side-on projections of these p.o.
With black are the stable and with orange colour the complex unstable ones. The
dashed arrows indicate that the wings of the {\sf X} are formed along the z
maxima of the orbits. The panels surrounding the central frame give the side-on
views of the non-periodic orbits used in the profile of Fig.~\ref{prof}c. The
numbers in the lower right corner of these panels corresponds to the energies
numbered along the $\Delta$ curve. We observe that smaller $\Delta$ values
correlate with more diffused orbits in phase space at the corresponding energy.
Orbit ``3'' is a particular case (see text).}
\label{noperorbs} 
\end{figure*}

In the case of complex instability the quantity $\Delta$ (Eq.~\ref{eq:delta}) is
$\Delta < 0$. We have observed that in our model, for a constant deviation from
the initial conditions of the periodic orbit, smaller values of $\Delta$
correspond to orbits more extended in configuration space. Fig.~\ref{noperorbs}
shows the individual orbits used to construct the profile of Fig.~\ref{prof}c.
In the central panel of Fig.~\ref{noperorbs} we plot $\Delta$ as \ej varies.
From ``1'' to ``8'' we mark in this diagram with heavy dots the locations of the
orbits we used. The $(y,z)$ projections of the corresponding p.o. are given in
the embedded frame. With black we have drawn the stable ones, while the orange
coloured are complex unstable ($\Delta$). The arrows help us understand that the
wings of the {\sf X} feature are formed along the $|z|$ maxima of the successive
orbits. In the frames around the central figure, the orbits numbered ``1'',
``7'' and ``8'' are quasi-periodic and have morphologies resembling those of the
x1v1 p.o. Counterintuitively, we observe at some energies that also chaotic
orbits in the neighbourhood of complex unstable x1v1 p.o. resemble
``$\frown$''-shaped structures (orbits numbered ``2'', ``4'', ``5'', ``6'').
However, they are much less confined in the configuration space. The orbit in
position ``3'' for \ej $= -0.37$ is a particular case, since it is more affected
by the presence of the family x1v2, which for \ej $=-0.37$ is simple unstable.
We will refer to it in detail below (Sec.~\ref{sec:x1m2v1}). In
Fig.~\ref{noperorbs} we observe that the most extended orbit in the
configuration space is orbit ``5'', which has the smallest $\Delta < 0$. The
same happens with orbits perturbed by $\Delta z$ and $\Delta x$ from x1v1 (not
shown in Fig.~\ref{noperorbs}).  

The explanation for the orbital behaviour  of the ``$\Delta$" orbits in
Fig.~\ref{prof} is again related to the phenomenon of stickiness. Stickiness
associated with complex instability has been recently studied extensively by
\cite{kpc11}. We find in our barred model perfect agreement with the results of
that study. The consequents in the 3D projections of the 4D space of section in
the neighbourhood of the complex unstable p.o. form spirals \citep{cfpp94,
pcp95}. In the 4D representation by means of the colour-rotation method we
realise that there is a smooth colour variation along the arms of these spirals.
For small perturbations of the initial conditions objects that are called
confined tori are formed in the neighbourhood of the ``$\Delta$'' p.o. in the 3D
projections of the spaces of section \citep{pf85a, pf85b, jo04, oea04}. Like in
\cite{kpc11} we find also in the model of the present paper that these are 4D
objects with an internal spiral structure and with a smooth colour variation on
their surfaces. For explaining the peanut and the {\sf X} structure, we are
interested in orbits that give a few tens of consequents on the 4D surfaces of
section. These are consequents that describe a first set of spirals on the
confined tori. Especially the first 10-20 consequents are located in parts of
the spirals very close to the p.o. 
During this time the orbital behaviour in the neighbourhood of the ``$\Delta$''
p.o. is characterised by stickiness \citep{kpc11}. This results to morphologies
having a certain similarity to quasi-periodic orbits around a stable orbit. This
explains the morphologies of the orbits ``2'',``4'',``5'' and ``6'' in
Fig.~\ref{noperorbs}. Diffusion in phase space happens after completing
thousands of consequents. However, for larger perturbations or for larger
energies of the orbits, diffusion happens earlier. 

\section{``OX'' vs ``CX'' profiles}
\label{sec:x1v2}

The side-on profiles given in Fig.~\ref{prof} resemble the kind of {\sf X}
features encountered in b/p bulges, which have wings that do not cross the
centre of the galaxy. A typical, example of this kind of b/p bulges is the
galaxy NGC~4710 (for a high-resolution image of this object, see the Space
Telescope Science Institute's news release STScI-2009-30). This is the kind of
{\sf X} features labelled by \cite{ar06} and \cite{betal06} as ``OX''
(off-centred). Indeed, in many images of galaxies the direction of the wings of
the {\sf X}, as revealed after applying an unsharp-masking filter on them,
clearly shows that their extensions towards the centre of the galaxy will not
cross it \cite[see also figure 2 in][]{px06}. However, in some other cases we
have a genuine {\sf X} feature, as is for example the edge-on galaxy HCG87A in
the Hickson 87 group \citep{hcks93} (for a high resolution image, see the
STScI-1999-31a news release). \cite{ar06} and \cite{betal06} call this type of
{\sf X}, ``CX'' (centred). The side-on ``OX'' profiles in Fig.~\ref{prof} are
based on orbits in the neighbourhood of x1v1 p.o. Possible  projection effects,
that could bring the wings of the ``x1v1-{\sf X}'' closer and form  ``CX''
instead of `OX'' profiles will be discussed in section \ref{sec:geom}. 
Below we examine the possibility of building true ``CX'' profiles using orbits.

\subsection{Possible x1v2 contribution}

In nearly side-on views of galaxies, true {\sf X}'s can be formed only with the
help of orbits associated with other families than x1v1. Orbits belonging to
the x1-tree, bifurcated closer to corotation, with the appropriate shape may
have in some models stable parts. Such an example is given by \cite{psa02}
for their standard model, based on the x1v4 family. Families bifurcated closer
to corotation, if populated, will build b/p bulges in which 
the ratio of the length of the bar to the length of the b/p distortion
is slightly larger than 1. However, in the present study we are looking for
b/p structures in which this ratio is $2.7\pm0.3$, as we mentioned in the
introduction.
 
\begin{figure}
\begin{center}
\resizebox{70mm}{!}{\includegraphics[angle=0]{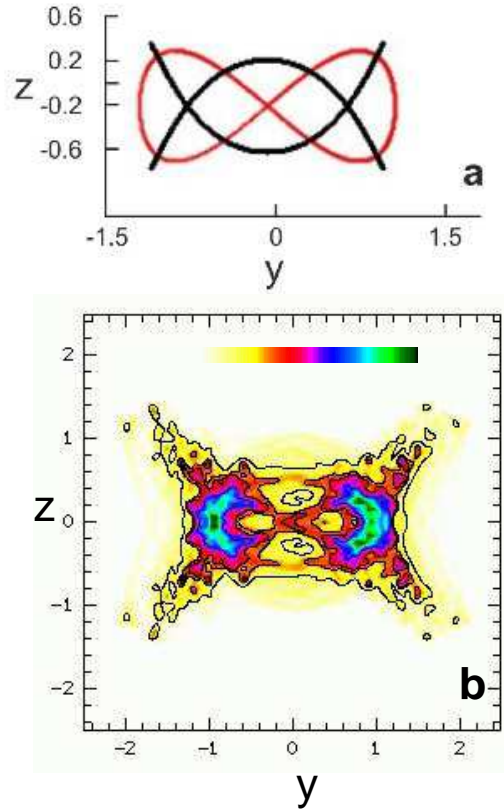}}
\end{center}
\caption{(a) Typical side-on profiles of x1v2 (red) and x1v1 and x1v1$^{\prime}$
(black) p.o. (b) A side-on profile similar to the one in Fig.~\ref{prof}b in
which we have substituted the orbits at \ej $= -0.41, -0.39$ and $-0.37$ with
non-periodic sticky orbits in the neighbourhood of x1v2. }
\label{x1v2prof} 
\end{figure}

Under those considerations, a family candidate to play this role would be x1v2,
bifurcated from x1 at a U$\rightarrow$S transition, in our model at
\ej\!\!$\approx -0.425$ (see Fig.~\ref{stabdiag}). The side-on projections of
the p.o. of this family have an $\infty$-type morphology \citep{spa02a}. In
Fig~\ref{x1v2prof}a we plot a typical side-on view of a x1v2 p.o. orbit (red)
together with the side-on projections of the p.o. x1v1 and x1v1$^{\prime}$.
Obviously the x1v2 orbits have qualitatively the appropriate shape to support a
central $\infty$-type feature in the $(y,z)$ plane. However, the x1v2 family is
initially simple unstable, inheriting the simple unstable character of x1, while
for larger \ej it evolves to double unstable (Fig.~\ref{stabdiag}a). Thus, the
mechanism of trapping quasi-periodic orbits around x1v2 p.o. cannot be applied. 

Nevertheless we may look for non-periodic orbits that will fill the empty
central parts in the profiles of Fig.~\ref{prof} converting them from ``OX'' to
``CX''-type. The range of energies that has to be covered is $-0.425 <$ \ej $<
-0.35$, i.e. roughly the interval within which x1v2 is simple unstable. When
this family becomes ``DU'' we do not expect any contribution to an observed
structure, since double instability leads to strong chaoticity \citep{kpc13}.  

We know \citep{kpc13} that in a range of \ej away from the immediate
neighbourhood of the bifurcating point\footnote{Tiny perturbations close to the
bifurcating point, at which x1v2 is introduced in the system, secure almost
regular behaviour for several Hubble times \citep{kpc13}. Nonetheless, this is
not of practical use because of the needed proximity of the initial conditions
of the perturbed orbits to the periodic orbit.} the unstable asymptotic curves
of the (U) x1v2 p.o. wind around  the rotational tori of the stable x1v1 and
x1v1$^{\prime}$ families that exist at the same energy. The construction of the
asymptotic curves is extensively described in \cite{kpc13}. Here, we will
examine how  this applies in the case of our 3D bar model. 

For energies beyond the bifurcation point, 
we have integrated x1v2 orbits perturbed by $\Delta x, \Delta p_x, \Delta z,$ or
$\Delta p_z$ for equal time as for the orbits composing the profiles of
Fig.~\ref{prof}. We first realised that such orbits, with \ej $< -0.35$, have
the appropriate sizes to fill the central empty regions of the profiles in
Fig.~\ref{prof}. We also found that a $\infty$-type morphology is supported for
the longest time by orbits perturbed by $\Delta x$. 
Perturbed x1v2 orbits by  $\Delta x \lessapprox 0.1$ are able to reinforce to
some degree an $\infty$-type morphology in their side-on projections. For
example, by substituting the orbits used for \ej $= -0.41, -0.39$ and $-0.37$ in
Fig.~\ref{prof}b by x1v2 orbits perturbed by $\Delta x = 0.1, 0.1$ and 0.05
respectively, we obtain the profile given in Fig.~\ref{x1v2prof}b. Despite a
small asymmetry with respect to the y=0 axis we can observe a ``CX'' central
feature.

\begin{figure}
\begin{center}
\resizebox{70mm}{!}{\includegraphics[angle=0]{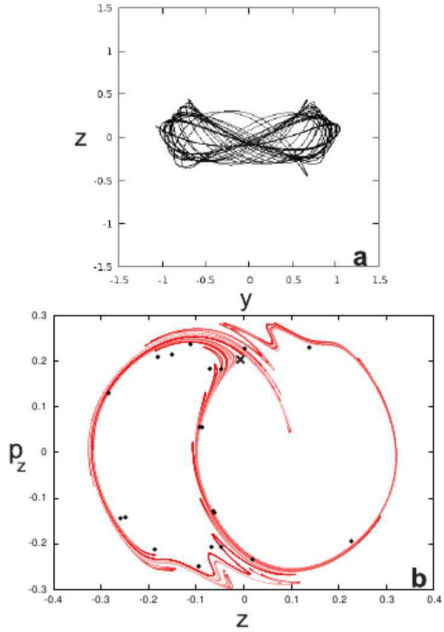}}
\end{center}
\caption{ (a) The $(z,y)$ projection of a chaotic orbit in the neighbourhood of
x1v2 at \ej$=-0.41$. Its initial conditions differ from the (U) p.o. by $\Delta
x = 0.1$. (b) A $(z,p_z)$ projection of the unstable manifold (red curve),
together with the first 18 consequents of the orbit in (a) (black points). }
\label{manx1v2} 
\end{figure}

By varying \ej\!\!, we realise that the ability of perturbed x1v2 orbits to
support an $\infty$-type side-on profile depends in general mostly on the
overall structure of the phase space in the region close to the p.o. Critical
for this purpose is the relative location of the x1v1 and x1v1$^{\prime}$
rotational tori with respect to the x1v2 p.o. For energies close beyond the
U$\rightarrow$S transition of x1 (\ej$\approx$ -0.425) the x1v1 tori are still
in the neighbourhood of the p.o. x1v2. At \ej = -0.42398, the stability index of
x1v2, which is associated with vertical perturbations, is $b_2=-2.2$. Its
initial
conditions are $(x_0,z_0,p_{x_0},p_{z_0})\approx(0.10476,0,0,0.055)$, while the
corresponding initial conditions for x1v1 at the same energy are $(0.10376,
0.159,0,0)$. Despite the fact that the orbits explore a 4D space, the proximity
of the x1v1 tori to the location of the x1v2 p.o. guides the asymptotic curves
of the x1v2 manifolds. As long as the unstable manifold of the U p.o. winds
around the x1v1 tori, the chaotic orbits follow a fuzzy x1v2 morphology. By this
we mean that during the time of integration the orbit has conspicuous x1v2
morphological elements for considerable time intervals. For example, the orbit
of Fig.~\ref{manx1v2}a has a hybrid morphology with features mainly from x1v2
and x1v1$^{\prime}$. We consider as hybrid orbits supporting a ``CX'' profile,
orbits that create a clearly discernible intersection point close to the
equatorial plane, i.e. at $|z| \lesssim 0.1$, as is the orbit in
Fig.~\ref{manx1v2}a. We find that about the first 20 consequents of such orbits
in the $(z,p_z)$ plane are projected very close to the unstable manifold of
x1v2. For the orbit of Fig.~\ref{manx1v2}a this can be seen in
Fig.~\ref{manx1v2}b, where we give the $(z,p_z)$ projection of the unstable
manifold (red curve) of x1v2 and the corresponding projections of the first 18
consequents of the orbit (black points). The location of the x1v2 p.o. is marked
with an $\times$ symbol. The red curve corresponds to the part  of the manifold
that winds around the x1v1 and x1v1$^{\prime}$ tori.

As the energy increases the distance between the p.o. x1v2 and
that of the x1v1, x1v1$^{\prime}$ tori increases and the diffusion of the
chaotic orbits from the neighbourhood of x1v2 into the available phase space is
fast \citep{kpc13}. 
However, we find anew an enhancement of x1v2 type morphologies at energies
$-0.39\lessapprox$ \ej $\lessapprox -0.37$. The new element that has to be taken
into account in this energy interval is the introduction in the system of the
planar stable family o1. This family is introduced in the system together with
its symmetric with respect to the y-axis family, o2 \citep{spa02b}. In
Fig.~\ref{stabdiag} we can see how o1 is introduced in the system at \ej
$\approx -0.39$, when the $b_1$ stability index of x1 becomes $b_1 < -2$.
3D chaotic orbits, sticky to o1, o2 tori have morphologies resembling that of 
 Fig.~\ref{manx1v2}a. 

Our analysis shows that there are orbits starting in the neighbourhood of x1v2
that can contribute to the formation of a ``CX''-type profile in the side-on
views of the models. We find them being sticky to tori of stable families
co-existing at the same energy. Then these orbits can be reached also by
perturbing the initial conditions of the stable orbits themselves, like the one
labelled with ``3'' in Fig.~\ref{noperorbs}, which we found by perturbing the
x1v1 initial conditions. However, the orbits reinforcing a ``CX''-type profile
do not share common features in the two other projections of the configuration
space. This underlines  the chaotic character of the orbits.

\subsection{The relative importance of non-periodic orbits and the x1mul2
family}
\label{sec:x1m2v1}
\begin{figure*}
\begin{center}
\resizebox{160mm}{!}{\includegraphics[angle=0]{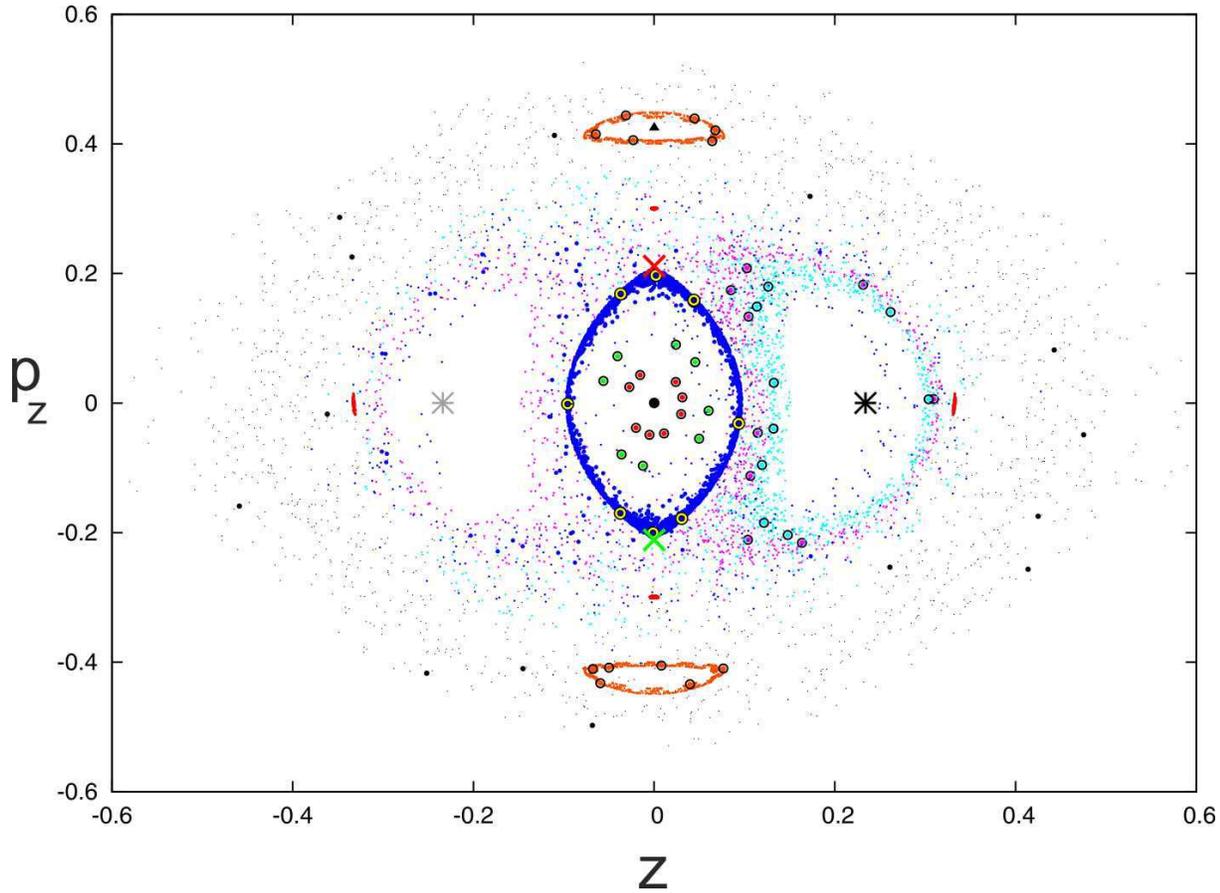}}
\end{center}
\caption{The $(z,p_z)$ projection of the 4D phase space of x1 orbits perturbed
in the $p_z$ direction, at \ej$=-0.41$. We indicate with black and grey ``$*$''
symbols the location of the x1v1 and x1v1$^{\prime}$ p.o. respectively. Red and
green ``$\times$'' symbols indicate the two branches of x1v2. The
``$\blacktriangle$'' at $(z,p_z)=(0.1,0.42)$ marks the location of the p.o.
x1mul2. Consequents corresponding to parts of the orbits that are plotted in
other figures are marked with $\odot$ symbols.}
\label{zpz} 
\end{figure*}

Trying to assess the significance of the presence in phase-space of each of
the families of p.o. for enhancing a ``CX'' side-on profile, we investigated
perturbations of their common ``mother family'', x1, away of the equatorial
plane . For this purpose we considered first x1 orbits on the equatorial plane
and we explored the phase space in the vicinity of x1 p.o. perturbed by $\Delta
p_z$. This gave us insight about the motion of particles, which are following
initially a x1 flow and are kicked out of the galactic plane. A typical energy,
where we can follow the influence of all these families to the orbital dynamics
is \ej = $-0.41$. It is typical because it is away from bifurcating points and
the families of p.o. retain their main stability character. At \ej = $-0.41$
coexist the families x1 (S), x1v1 (S) and x1v2 (U) (Fig.~\ref{stabdiag}). For
describing the properties of these orbits which are relevant to the side-on
morphology of the peanut structure, we base our description on the $(z,p_z)$
projection of the 4D phase space. This is given in Fig.~\ref{zpz}. In this
figure the x1 p.o. is located at (0,0) marked with a heavy black dot.

\begin{figure}
\begin{center}
\resizebox{80mm}{!}{\includegraphics[angle=0]{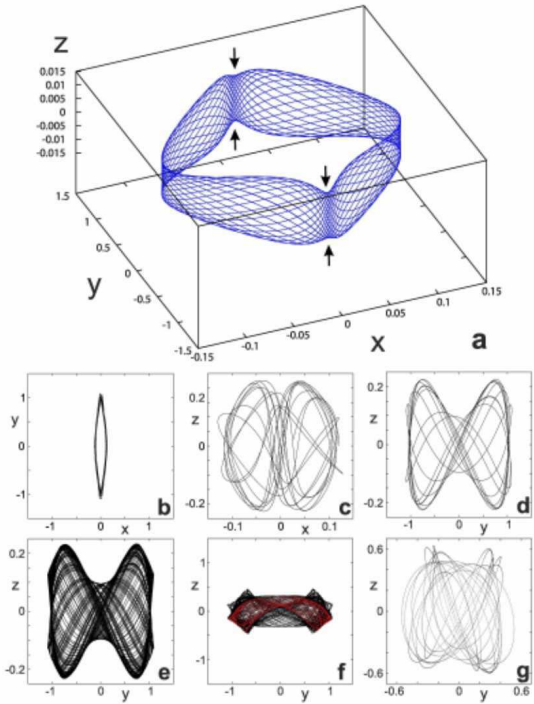}}
\end{center}
\caption{The evolution of the morphology of quasi-periodic orbits around x1, as
we approach the initial conditions of x1v2 at \ej$=-0.41$. (a) A typical
quasi-periodic orbit corresponding to the green invariant curve around x1 in
Fig.~\ref{zpz}. (b,c,d) The sticky orbit that forms the dark blue ring in 
Fig.~\ref{zpz} after 10 x1 periods. (e) The (y,z) projection of the same orbit
after 2249 intersections with the surface of section. It still has an
``$\infty$''-type morphology. (f) The side-on view of the orbit coloured light
blue in Fig.~\ref{zpz}, which is sticky to the x1v1 and
x1v1$^{\prime}$ tori. The overplotted red part corresponds to 10 x1 periods. (g)
The (y,z) projection of the x1 orbit perturbed by $p_z$=0.5. It has a chaotic
character. Note the different scales of the axes on the panels. }
\label{zpzorb} 
\end{figure}

At \ej$=-0.41$ small perturbations of x1 by $\Delta p_z$ bring in the system
quasi-periodic orbits of cylindrical shape. They are characterised by a local
$|z|$ minimum for $y=0$ in the $(y,z)$ projections, similar to that of
quasi-periodic orbits around x1 (Fig.~\ref{thickx1}a). We notice though, that at
this \ej we have a second, local,  $|z|$ minimum at $x=0$, best observed in the
$(x,z)$ projection. This means that we have two characteristic sets of dents
along the boundaries of the quasi-periodic orbits. Such an orbit is depicted in
Fig.~\ref{zpzorb}a and the local minima under discussion for $x=0$, are marked
with arrows. The innermost two elliptical curves around $(z,p_z)=(0,0)$ in
Fig.~\ref{zpz} correspond to this kind of orbits. The $\odot$ symbols along this
and other orbits in Fig.~\ref{zpz} correspond to the consequents of parts of the
orbits that are used for presenting their morphologies (see below).  

As we increase the $\Delta p_z$ perturbation, always for constant \ej$=-0.41$,
both local minima deepen, approaching $z=0$. The increase of the $\Delta p_z$
perturbation brings us closer to the initial conditions of x1v2, which is marked
in Fig.~\ref{zpz} with a red ``$\times$'' symbol close to (0,0.2). With a green
``$\times$'' symbol close to $(0,-0.2)$ we mark the x1v2$^{\prime}$ orbit.
Essentially these two orbits are two branches of the same family \citep{spa02a}.
We find that starting from x1 and approaching the
initial conditions of x1v2 by increasing $p_z$, the boundaries of the
non-periodic orbits become more and more similar to the shape of the x1v2 p.o.
This is 
similar to what was happening as we were approaching the
bifurcation point of x1v1 by increasing the energy in Section~\ref{sec:start}.

By reaching the perturbed by $p_z = 0.195$ x1 orbit we encounter a new kind of
orbital behaviour. The first 1500 consequents form an elliptical, lemon-shaped,
ring in the $(z,p_z)$ projection, plotted with dark blue colour in
Fig.~\ref{zpz}. Then the projected consequents drift to the inner part of the
ring and the ring thickens. After 2249 intersections with the surface of
section, the consequents depart from the area of the ring and fill a region
around two ``empty'' lobes in the left and the right side of Fig.~\ref{zpz}. In
the interior of these two lobes are projected the initial conditions of x1v1
(right) and x1v1$^{\prime}$ (left), marked with black and grey ``*'' symbols
respectively. The first 50 consequents of the perturbed orbit after their
departure from the ring area are plotted with intermediate sized blue dots in
Fig.~\ref{zpz} and can be seen mainly around the left lobe. The morphology of
this orbit within 10 periods of x1 (corresponding to the consequents marked with
yellow ``$\odot$'' symbols on the blue ring) is given in Figs.~\ref{zpzorb}b,c
and d for the $(x,y)$, $(x,z)$ and $(y,z)$ projections respectively. There is a
conspicuous enhancement of an $\infty$-type morphology in the $(y,z)$ view. If
we consider the orbit for the whole time it remains on the blue ring, this
morphology persists. This can be seen in its $(y,z)$ projection given in
Fig.~\ref{zpzorb}e.  We underline also the narrowness of the orbit along the $x$
axis. This may be crucial for the appearance of the ``CX'' morphology in the
edge-on views of disc galaxies, since it secures a small distance between the
front and the back branches of the $\infty$- shaped orbit.

By increasing further the perturbation in the $p_z$ direction, we come very
close to the initial conditions of x1v2. However, by varying only the $p_z$
coordinate we will never reach the x1v2 p.o. since we keep constant the $x_0$
initial coordinate, equal with the $x_0$ value of x1. This is smaller than the
x1v2 $x_0$ initial condition by about 0.011. Orbits in the neighbourhood of the
p.o. x1v2 recede away from it following trajectories in the direction specified
by its unstable maniflods (Fig.~\ref{manx1v2}b). The ``empty'' lobes of
Fig.~\ref{zpz} are like those given in Fig.~\ref{manx1v2}b. For example, small
displacements from x1v2 (red $\times$) \textit{in the $z$ direction} lead to
clouds of points around the lobes in Fig.~\ref{zpz}. The light blue and magenta
coloured points belong to orbits perturbed from x1 by $(\Delta z,\Delta
p_z)=(0.15,0.2)$ and $(0.2,0.2)$ respectively. The empty in Fig.~\ref{zpz}
regions are occupied by tori around x1v1 and x1v1$^{\prime}$ and are needed
thousands of consequents to fill them in the $(z,p_z)$ projection. In
Fig.~\ref{zpzorb}f we give the $(y,z)$ projection of the first of these two
orbits integrated for 100 x1 periods. On top of it we colour with red the part
that corresponds to 10 x1 periods. The consequents of this part of the orbit are
given with light blue ``$\odot$'' symbols around the right ``empty'' lobe in
Fig.~\ref{zpz}. This additional information of the shape of the orbits within
realistic times is necessary in order to understand the morphologies they
support. Nevertheless, for understanding the dynamical phenomena behind the
observed orbital dynamics we frequently need to extend our calculations to times
of the order of a Hubble time or more. 

Orbits of x1 perturbed by $p_z$ larger than the $p_{z_0}$ x1v2 initial condition
fill roughly a ring surrounding the area defined by the initial conditions of
x1v2 and x1v2$^{\prime}$ and the ``empty'' lobes to the right of x1v1 and to the
left of x1v1$^{\prime}$ in Fig.~\ref{zpz}. In this area exist four tiny
stability islands. Starting close to $(z,p_z)\approx(0,0.3)$ we find them in
azimuthal distances of about $90^{\circ}$ (coloured red). These islands are the
projections of rotational tori around a stable p.o. of multiplicity four. This
orbit enhances the boxiness of the peanut mainly by forming a zone of stickiness
connecting the four tori. Orbits with an ab initio strong chaotic
behaviour are found for $p_z \gtrapprox 0.5$. The morphology of the perturbed by
$p_z =0.5$ x1 orbit is already  quite chaotic in the $(y,z)$ projection within
10 x1 periods, as can be observed in Fig.~\ref{zpzorb}g. The consequents
corresponding to the part of the orbit depicted in Fig.~\ref{zpzorb}g are given
with heavy, this time black, dots in Fig.~\ref{zpz}. They are distributed in a
zone that surrounds the orbits sticky to x1v1 and x1v2. The same holds also for
the next 1000 consequents (drawn with lighter black dots). We observe however,
that there is no substantial mixing of the outer chaotic zone, which includes
the black dots, with the inner structure containing the lemon-shaped ring, the
x1v2 and x1v2$^{\prime}$ p.o. and the ``empty'' lobes around them. Stickiness is
ubiquitous in Fig.~\ref{zpz}. We practically reach the same topology of phase
space by perturbing the x1 orbits in the $p_z$ direction and by perturbing the
x1v2 orbit itself. One can compare Fig.~\ref{zpz} with Fig.~\ref{manx1v2}b but
also with figure 5 in \citet{kpc13}. The basic feature that structures the phase
space is the presence of the area occupied by the x1v1 and x1v1$^{\prime}$ tori.
This is a typical property of the 4D spaces of section of our system. 

\begin{figure}
\begin{center}
\resizebox{82mm}{!}{\includegraphics[angle=0]{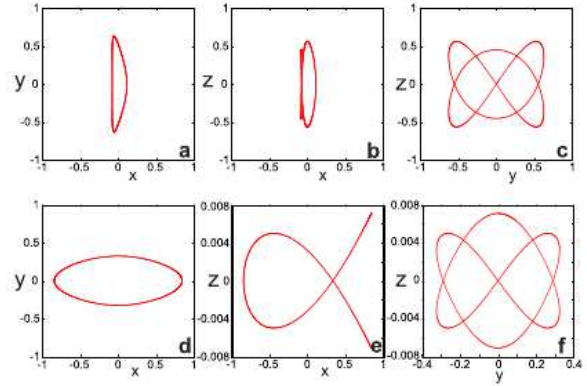}}
\end{center}
\caption{(a,b,c) The $(x,y), (x,z)$ and $(y,z)$ projections of the x1mul2
family. (d),(e),(f) the corresponding projections of the x2 family. Both are
given at \ej$=-0.3976$. Note the difference in the extent of the two orbits in
the $z$ direction.}
\label{mul2orbs} 
\end{figure}

\begin{figure}
\begin{center}
\resizebox{80mm}{!}{\includegraphics[angle=0]{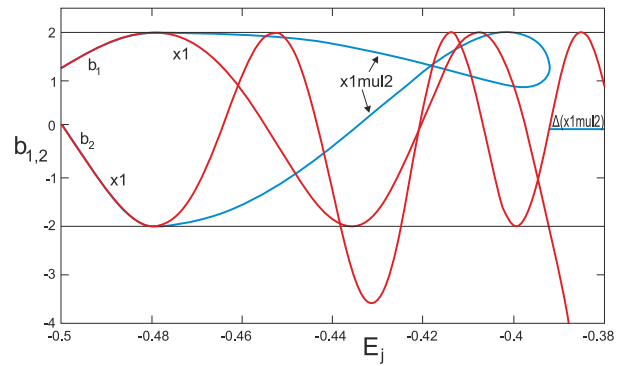}}
\end{center}
\caption{The red curves give the evolution of the stability indices of the x1
family when it is considered being of multiplicity 2. The family x1mul2 is
introduced in the system at \ej $\approx -0.479$.}
\label{mul2stab} 
\end{figure}

\begin{figure}
\begin{center}
\resizebox{88mm}{!}{\includegraphics[angle=0]{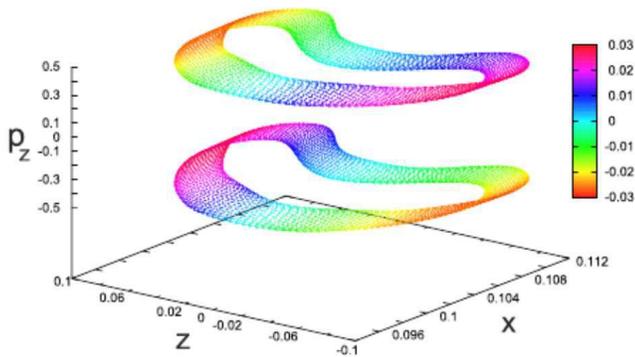}}
\end{center}
\caption{4D rotational tori around x1mul2 at \ej=$-0.41$. The consequents are
coloured according to their $p_x$ values (colour bar at the right side of the
figure).}
\label{tworotori} 
\end{figure}

We have left for the end the description of the two tori that appear centred
around $(z,p_z)=(0.1,\pm 0.42)$ in Fig.~\ref{zpz}, because they bring in the
system an important new family, which has to be discussed separately. Both tori
belong to the same family, i.e. it is a p.o. of multiplicity 2. Just the size of
the projected tori indicates the
importance of this family. Its shape can be seen in the panels of the upper row
of Fig.~\ref{mul2orbs} (a,b,c). It is quite similar to the family x2mul2 found
by \cite{spa02a} to be bifurcated from x2 (panels Fig.~\ref{mul2orbs}d,e,f). Its
$(y,z)$ view has a shape that contributes to the
enhancement of a ``CX'' profile. In accordance with x2mul2 we call this family
x1mul2. Both p.o. in Fig.~\ref{mul2orbs} are given at \ej$=-0.3976$ shortly
after the bifurcation of x2mul2 from x2. We note that the vertical extent of
x1mul2 is more than 60 times larger than that of x2mul2, hence it is a much more
significant family for the overall morphology of the boxy bulges. The x1mul2
family is a vertical bifurcation of x1 at \ej $\approx -0.479$. In order to
describe the connection between x1 and x1mul2 we give in Fig.~\ref{mul2stab} the
evolution of the stability indices of the x1 family when it is considered being
of multiplicity 2 (red curves). The stability indices of x1mul2 emerge  at \ej
$\approx -0.479$, where the $b_2$ index becomes tangent with the $b=-2$ axis. We
observe that x1mul2 is introduced in the system before the x1v1 and x1v2
families and extends to a large energy interval. In the interval $-0.48
\lessapprox$ \ej $\lessapprox -0.392$ it is stable, while for \mbox{\ej
$\gtrapprox -0.392$} it becomes complex unstable. We note that the non-zero
coordinates of the initial conditions of the p.o. of the x1mul2 family are $x_0$
and $p_z$ like the initial conditions of x1v2.
The structure of the 4D space of section in the neighbourhood of x1mul2 is as
foreseen by \citet{kpp11}. Namely we have a couple of rotational tori with
smooth colour variation along them. A typical example is given in
Fig.~\ref{tworotori}. It depicts two 4D rotational tori around x1mul2 at
\ej=$-0.41$. The chosen 3D projection is $(x,z,p_z)$, while the consequents are
coloured according to their $p_x$ values.

\begin{figure}
\begin{center}
\resizebox{80mm}{!}{\includegraphics[angle=0]{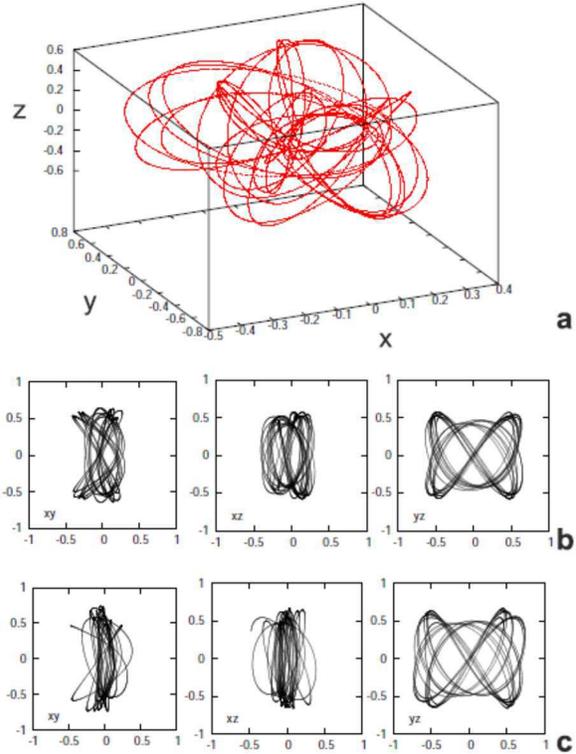}}
\end{center}
\caption{(a) A quasi-periodic orbit associated with x1mul2 at \ej$=-0.4$ viewed
from a line of sight specified by the angles $(\theta, \phi)$=(60\dgr,335\dgr).
(b) (from left to right) The $(x,y), (x,z)$ and $(y,z)$ projections of the same
orbit, show that it supports an  ``$\infty$'' type profile. (c) An orbit in the
immediate neighbourhood of a complex unstable x1mul2 p.o. at \ej$=-0.39$.}
\label{np_mul2orb} 
\end{figure}

Side-on profiles of quasi-periodic and sticky orbits around x1mul2 support boxy
structures. On top of this the orbits are self-intersecting close to
$(y,z)=(0,0)$ in their side-on views. Most of the quasi-periodic orbits can be
thought as thicker versions of the orbit depicted in the upper row of
Fig.~\ref{mul2orbs}. 
Nevertheless, in other projections these orbits may have a complicated or
chaotic looking morphology. As an example we give in Fig.~\ref{np_mul2orb}a,b
the quasi-periodic orbit that deviates from x1mul2 by $\Delta x = -0.111$,
integrated for 12 periods of x1v2 at \ej = $-0.4$. In Fig.~\ref{np_mul2orb}a the
projection 
particular structure. However, in Fig.~\ref{np_mul2orb}b, right panel, we have a
conspicuous support of a boxy feature with an ``CX'' morphology. The face-on
view $(x,y)$, left panel of Fig.~\ref{np_mul2orb}b, does not have a clear x1,
character as in Fig.~\ref{mul2orbs}a, or Fig.~\ref{zpzorb}b at the same energy.
It has a morphology resembling quasi-periodic orbits around 3:1 stable orbits on
the equatorial plane. Even so, it retains its narrowness in the $x$ direction,
especially close to $y=0$. 
Similar conclusions we draw by observing orbits in the immediate neighbourhood
of x1mul2 close after the S$\rightarrow\Delta$ transition. An example is given
in Fig.~\ref{np_mul2orb}c for the orbit that deviates by $\Delta p_z=0.2$ from
the p.o at \ej$=-0.39$. The orbit in Fig.~\ref{np_mul2orb}c shares the same
properties with the one in Fig.~\ref{np_mul2orb}b in supporting a ``CX' side-on
profile. 

\subsection{The phase-space close to complex unstable p.o. of
multiplicity 2}
In previous papers the structure of phase space has been studied in the
neighbourhood of complex unstable \textit{simple} periodic orbits, but not close
to ``$\Delta$'' p.o. of multiplicity 2 or higher. Questions that arise concern
the shape of the `confined tori'' \citep{pf85a, pf85b} of multiplicity 2, their
internal spiral structure \citep{cfpp94, pcp95} and their 4D morphology
\citep{kpc11}. For this purpose we will briefly describe here the phase space
structure in the neighbourhood of ``$\Delta$'' x1mul2 orbits, which are of
multiplicity 2.
 
In the case of the simple periodic x1v1 family we know \citep{kpc11} that close
beyond a S~$\rightarrow$~$\Delta$ transition the 4D representations of the
confined tori are almost 2D, slightly warped, disky objects with a smooth colour
variation along their surfaces. Further away from the transition point the
confined tori obtain a toroidal structure and finally, beyond a critical energy,
we have scattered consequents in the 4D space. \citet{kpc11} describe also
phenomena of stickiness to confined tori.

\begin{figure*}
\begin{center}
\resizebox{170mm}{!}{\includegraphics[angle=0]{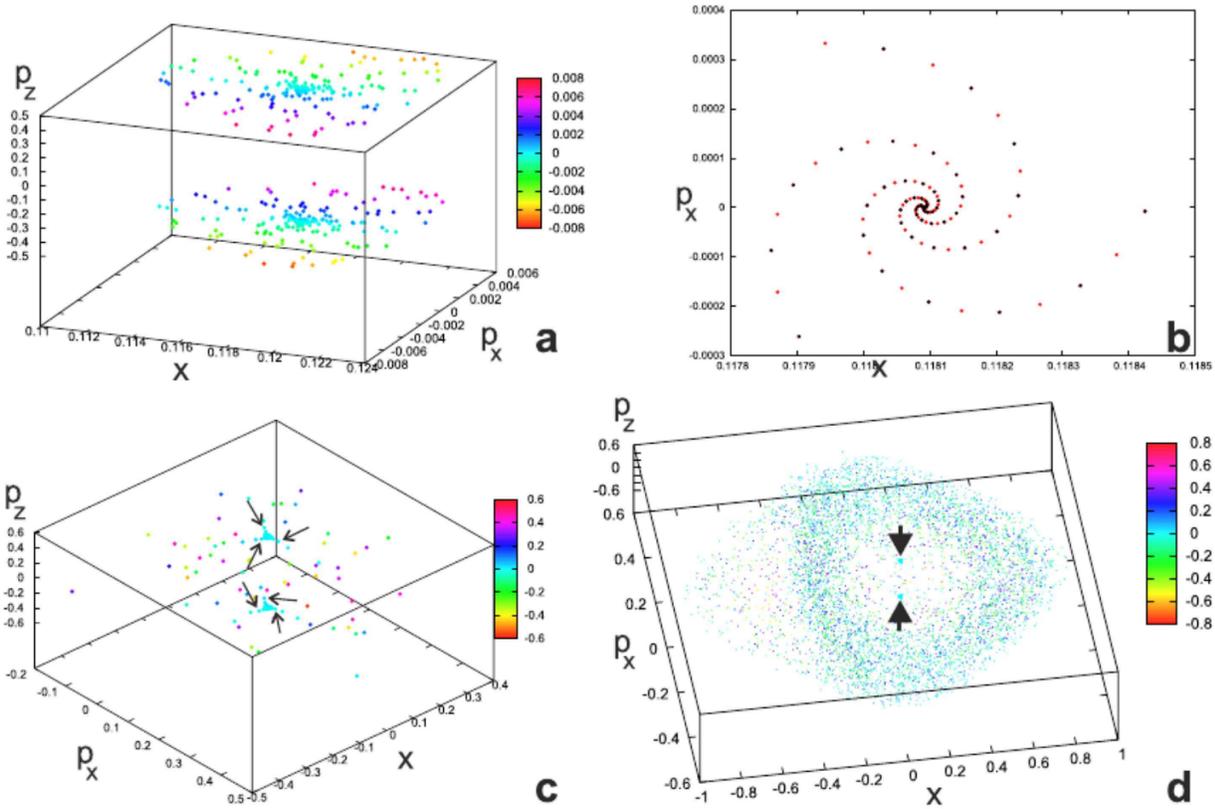}}
\end{center}
\caption{The evolution in phase space of a chaotic orbit in the immediate
neighbourhood of the complex unstable x1mul2 p.o. at \ej$=-0.392$. (a) 4D
visualisation of the simultaneous formation of an incomplete confined torus
(see text) of double
multiplicity after 400 intersections. (b) The spiral in the neighbourhood of the
x1mul2 p.o. Red dots belong to the upper and black to the
lower ``confined torus'' in (a). In (c) we observe how the consequents start
occupying a larger volume of the phase space after 470 intersections and in (d)
the last stage of the morphological evolution of the chaotic orbit after 2000
intersections. The
configuration does not change even after 8000 intersections. }
\label{compm2} 
\end{figure*}

This kind of behaviour is reproduced in the case of the x1mul2 family, however
with some peculiarities due to the double multiplicity of the p.o. Let us
consider an orbit at \ej$=-0.392$ that differs from the p.o. only by $\Delta x =
-10^{-5}$. We chose here such a small perturbation, because we want at this
point to describe in detail the underlying dynamics. We use for our description
the $(x,p_x,p_z)$ 3D projection and we colour the consequents according to their
$z$ values. After 400 intersections two disky objects tend to form centred at
$(x,p_x,p_z)\approx(0.118,0,\pm 0.47$), as we can see in Fig.~\ref{compm2}a.
There is no colour mixing as we move across these objects. We have rather a
smooth colour variation, meaning that all requirements exist for the formation
of a confined torus in 4D. The way the consequents tend to fill these two disky
structures follows a rule that can be better understood in the $(x,p_x)$
projection, i.e. if we observe the orbit of Fig.~\ref{compm2}a from above. We
give the first 100 consequents of the orbit in the $(x,p_x)$ projection in
Fig.~\ref{compm2}b. They form a set of 3 spirals with the p.o. in the centre of
the configuration, like the spirals in the complex unstable p.o. of single
multiplicity described in \citet{kpc11}. Three successive consequents belong to
three different spiral arms. However, along each one of the spiral arms the
successive consequents belong alternately to one of the two disky objects
depicted in Fig.~\ref{compm2}a. Red dots in Fig.~\ref{compm2}b belong to the
upper one in Fig.~\ref{compm2}a $(p_z>0)$, while black to the lower one
$(p_z<0)$. As we continue integrating the orbit, after the first 470
consequents, three tips are formed in each disky object. The subsequent
consequents initially elongate these three tips and then they diffuse in the 3D
phase space with their colours mixed (Fig.~\ref{compm2}c). The diffusion starts
before the two confined tori are filled by the consequents. The tips of the
three extensions of each uncompleted confined torus are indicated with arrows in
Fig.~\ref{compm2}c. The whole areas of the confined tori as depicted in
Fig.~\ref{compm2}a are included now within the triangular light blue areas out
of which the consequents diffuse in the 4D space. After 800 consequents we
clearly see that we observe a unique cloud of points. The final stage of this
evolution, reached after 2000 consequents is given in Fig.~\ref{compm2}d. The
situation remains like this, as we checked, even after 8000 consequents and is
characterised by the formation of a volume around the p.o in which the
distribution of consequents is sparse. The majority of the consequents is
trapped in a cloud of points surrounding this rather ``empty'' volume
(Fig.~\ref{compm2}d). For even larger integration times we finally have a
uniform distribution of consequents in all 3D projections. These results are of
interest for the study of autonomous Hamiltonian systems in general. However,
for the particular application we study here, we have to remember that we are
interested for the orbital dynamics within tens and not thousands of dynamical
times. Thus, we have to keep in mind that also close to multiplicity 2 complex
unstable p.o. the spirals in phase space keep the initial tens of consequents in
a small area of the phase space. Thus, they prevent the fast diffusion of orbits
starting close to the $\Delta$ p.o. and hence they reinforce, within this time,
structures associated with them.

We note that for \ej=$-0.392$, besides x1mul2 the x1v1 family is also complex
unstable. Perturbations of x1v1 by $\Delta x = 10^{-4}$ keep the consequents on
a disky confined torus for thousands of consequences. So, during long times we
do not have mixing in phase space due to the presence of two ``$\Delta$'' p.o.

\subsection{A new side-on profile}
Family x1mul2 is proven to be a very important family of p.o. for three reasons:
(i) it exists over a large energy range having extended stable parts. (ii) it
starts existing \textit{before} the \ej of the vertical 2:1 resonance, i.e. it
may trap material around it before x1v1. (iii) it builds a sticky
zone connecting its two rotational tori, which is located around the x1v1 and
x1v1$^{\prime}$ tori region (Fig.~\ref{zpz}). Then on one hand we have orbits
sticky to the x1mul2 tori and on the other hand the sticky zone blocks the easy
communication between the x1v1 - x1v1$^{\prime}$ tori region and the outer
parts of the phase space. 

\begin{figure}
\begin{center}
\resizebox{80mm}{!}{\includegraphics[angle=0]{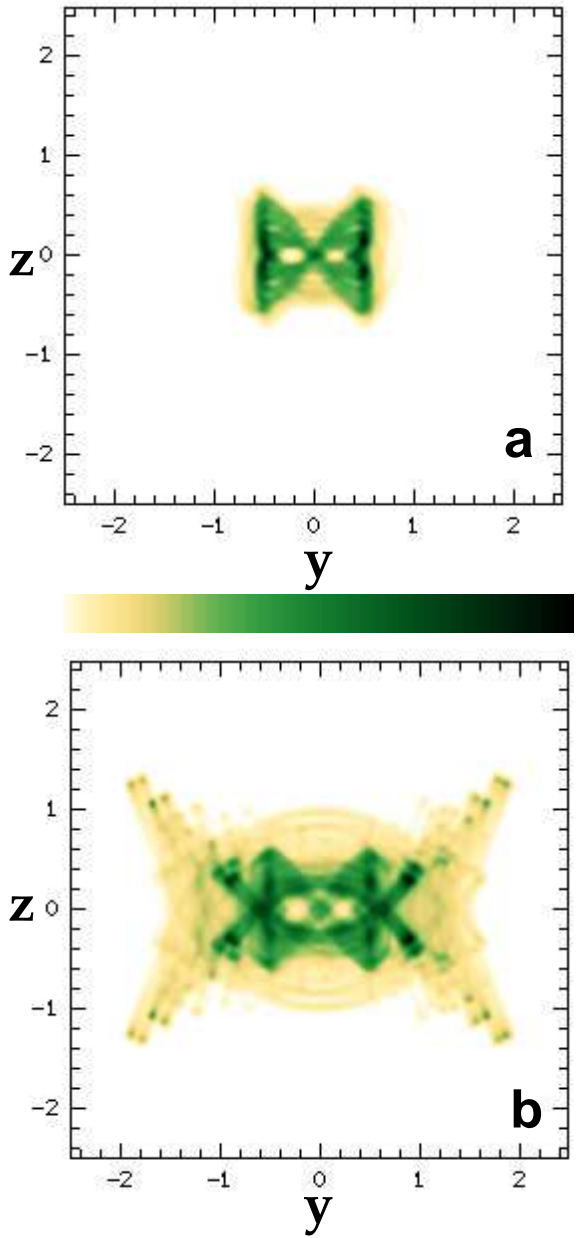}}
\end{center}
\caption{(a) A side-on profile constructed purely by orbits close to x1mul2 p.o.
(b) A composite side-on profile in which we have combined the profile in (a)
with the profile in Fig.~\ref{prof}c. Both profiles support a ``$\infty$''-type
feature.}
\label{multiprof} 
\end{figure}

Under these considerations our model indicates the possibility of b/p bulges
consisting partly, or totally, by orbits associated with the x1mul2 family. A
side-on profile composed by quasi-periodic and sticky orbits associated with the
orbits of x1mul2 is given in Fig.~\ref{multiprof}a. We have taken five perturbed
by $p_z \approx 0.025$ x1mul2 orbits for \ej $= -0.46, -0.44, -0.42, -0.40$ and
$-0.39$. The x1mul2 p.o. at these energies are stable except for \ej $= -0.39$,
which is complex unstable. They form a clear butterfly structure with dimensions
$(\Delta y \times \Delta z) \approx (1.6 \times 1.2)$, i.e. it reaches a height
$|z| \approx 0.6$. Including further orbits from the neighbourhood of x1mul2
orbits at higher energies results to a larger boxy bulge, however without
retaining the conspicuous butterfly morphology. Combined with orbits building an
{\sf X}-shaped bulge, the ``butterfly'' will be located in the central area of
the profile. How much pronounced it will appear in a model depends on the
coexistence of other families at the same regions and on how much each of them
will contribute to the building of the boxy bulge. For example the profile in
Fig.~\ref{multiprof}b consists of the orbits close to x1v1 p.o., which have
built the profile in Fig.~\ref{prof}c adding the orbits that give us the profile
in Fig.~\ref{multiprof}a. A central butterfly feature is in this case
discernible. In other models with with less extended ``$\Delta$'' parts in their
stability diagrams the role of x1mul2 is more pronounced.

\section{Geometrical considerations}
\label{sec:geom}
The projection angle under which we view the b/p bulges is expected to play a
role in their observed morphology. At any rate we study the morphology of disc
galaxies observed nearly edge-on, thus practically we want to investigate just
the changes
introduced by the rotation of the galaxy around its rotational axis. By rotating
the \textit{non-periodic} orbits that compose the profiles we have presented in
Fig.~\ref{prof}, Fig.~\ref{x1v2prof} and Fig.~\ref{multiprof} we first realise
that  non-periodic orbits in the neighbourhood of x1v2 and x1mul2 p.o.
contribute to the formation of a b/p or ``$\infty$'' feature only close to
their side-on views. Away from this projection angle they only contribute to the
formation of a boxy bulge, remaining confined in the $z$ direction. 

\begin{figure}
\begin{center}
\resizebox{80mm}{!}{\includegraphics[angle=0]{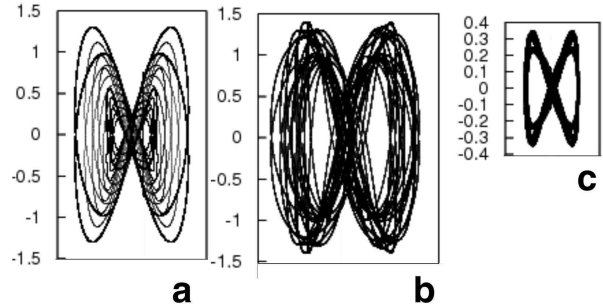}}
\end{center}
\caption{(a) A set of x1v1 and x1v1$^{\prime}$ orbits viewed at an angle $\theta
= 15^\circ$ (see text). (b) Orbit ``8'' of Fig.~ \ref{noperorbs} viewed
at the same projection. (c) Orbit ``1'' of Fig.~ \ref{noperorbs} projected with
$\theta = 8^\circ$. These are cases in which we have an ``$\infty$'' or
``{\sf X}'' feature away from the side-on views. The orbits in all three frames
are given in the same size.}
\label{geom} 
\end{figure}

As regards the orbits associated with the x1v1 family, besides the side-on view
that gives the ``OX'' profiles (Fig.~\ref{prof}), there is another projection
angle for which non-periodic orbits in the neighbourhood of x1v1 p.o. could
reinforce this time a ``CX'' type profile. For this, one has to consider both
representatives of x1v1 and x1v1$^{\prime}$ at each energy. These projection
angles are close to the end-on view of the bar, i.e. we have for the box a ratio
$r_{b/p}=R_{max}/z_{max} \lesssim 1$. In addition the resulting morphology is
closer to  an ``$\infty$''-type or peanut-shaped structure, than to an {\sf
X}-shaped feature with straight line segments as wings, as is for the side-on
views. In Fig.~\ref{geom}a we give a set of x1v1 and x1v1$^{\prime}$ p.o. in the
energy range $-0.41< $\ej $< -0.27$, from the point of view
$(\phi,\theta)$=(90$^\circ$,15$^\circ$) (for the end-on view we have
$(\phi,\theta)$=(90$^\circ$,0$^\circ$)).  Non-periodic orbits close to the x1v1
and x1v1$^{\prime}$ p.o. however, do not build b/p profiles with features as
sharp as in their side-on projections when combined and projected in other
projection angles. The ``$\infty$''-like structures they build are formed when
viewed in a range of $\theta$ angles centred around an angle for which the x1v1
\textit{p.o.} can reproduce the same feature much sharper. This angle is $\theta
\approx 15^\circ$ for quasi-periodic orbits after the complex unstable part of
x1v1 (Fig.~\ref{stabdiag}). For quasi-periodic orbits before the complex
unstable part of the family (\ej$ < -0.3965$) however, the ``$\infty$'' feature
is best viewed with $\theta \approx 8^\circ$. In Fig.~\ref{geom}b we give the
orbit ``8'' of Fig.~\ref{noperorbs} viewed with angles
$(\phi,\theta)$=(90$^\circ$,15$^\circ$) and in Fig.~\ref{geom}c the orbit ``1''
of Fig.~\ref{noperorbs} viewed with angles
$(\phi,\theta)$=(90$^\circ$,8$^\circ$). They are given together with their
symmetric orbits with respect to the equatorial plane.
An angle  $\theta \approx 15^\circ$ is proposed by \citet{eskoetal14} for the
viewing angle of the ``{\sf X}-shaped'' bulge of the Milky Way. A b/p structure
made by orbits associated with the x1v1 family projected at this angle is
compatible with our findings.

\begin{figure}
\begin{center}
\resizebox{80mm}{!}{\includegraphics[angle=0]{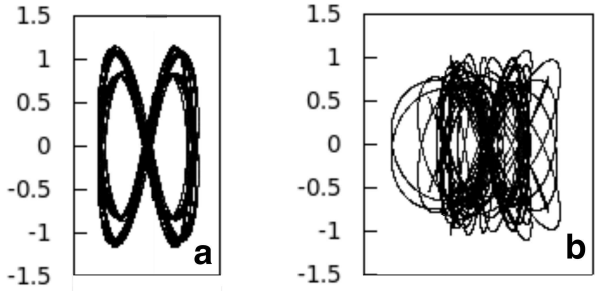}}
\end{center}
\caption{Comparison of a quasiperiodic and a chaotic orbit supporting 
an ``$\infty$''-type morphology close to their end-on views. In (a) we give
orbit ``7'' and in (b) figure ``6'' of Fig.~\ref{noperorbs}. In both cases we
have $\theta= 13^\circ$}
\label{geom2} 
\end{figure}

The projections away from the side-on view of the non-periodic orbits in
Fig.~\ref{noperorbs}, which are in the neighbourhood of ``$\Delta$'' x1v1 p.o.
(``3'' to ``6''), give evidently less pronounced b/p features, although they
have always a boxy character. In Fig.~\ref{geom2} we compare the morphology of
the orbits ``7'' and ``6'' of Fig.~\ref{noperorbs} when viewed with angles
$(\phi,\theta)$=(90$^\circ$,13$^\circ$). We plot them always together with their
symmetric with respect to the equatorial plane counterparts. The orbits in
Fig.~\ref{geom2}a with \ej$=-0.29$ are quasi-periodic, while those in
Fig.~\ref{geom2}b with \ej$=-0.31$ are chaotic. We realise the degree at which
they enhance the ``$\infty$''-type morphology close to their end-on views. We
note that if we consider all
the orbits of the sample of Fig.~\ref{noperorbs} at a nearly end-on view the
orbits with the larger energies are the most important for characterising the
overall morphology of the b/p bulge. Orbits at low energies remain in low
heights above the $z=0$ plane and thus they do not shape the outer parts of the
bulge.

\begin{figure}
\begin{center}
\resizebox{80mm}{!}{\includegraphics[angle=0]{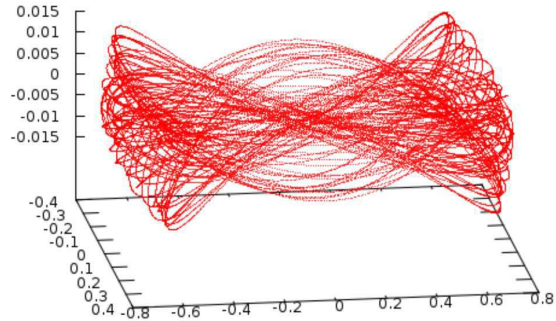}}
\end{center}
\caption{Individual orbits may create b/p structures away from edge-on views.
The orbit corresponds to the 4th invariant curve in Fig.~\ref{xxdot} perturbed
by $z_0 = 0.01$, starting at $p_x =0$. The projection angles are 
$(\phi,\theta)$=(60$^\circ$,82$^\circ$).} 
\label{projorb} 
\end{figure}

We note that one can in many cases find combinations of $(\phi,\theta)$
projection angles for individual non-periodic orbits in the neighbourhood of
x1v1 that projected resemble structures with a ``CX'' type morphology. As an
example we give the perturbed by $\Delta z = 0.01$ orbit corresponding to the
fourth invariant curve around x1 in Fig.~\ref{xxdot}, starting with $p_x=0$,
from a point of view defined by the angles (60$^\circ$,82$^\circ$)
(Fig.~\ref{projorb}). It is integrated for about 50 dynamical times of x1 at the
same energy (\ej$=-0.438225$). However, away from the edge-on orientation (in
our example in Fig.~\ref{projorb} $\phi=60^\circ$) the combination of many
orbits projected with the same angles in the cases we tried lead in the best
case to a boxy, but not to a b/p structure.

\section{Discussion and Conclusions}
\label{sec:concl}

The main result of the present study is that there is a significant contribution
from sticky chaotic orbits to the building of the body of b/p structures in the
side-on profiles of disc galaxy models. The families of the
x1-tree can be considered as the backbone of the peanut structure in two ways.
On one hand they trap around them quasi-periodic orbits and on the other hand
they create obstacles for chaotic orbits that hinder their diffusion in phase
space. These obstacles are the tori around orbits belonging to families of the
x1-tree (mainly of x1v1 and x1v1$^{\prime}$). The tori act as attractors for
chaotic orbits, which become sticky to them. The important region for the
peanut structure is the \ej interval of the radial-ILR and vertical-ILR. We find
that in the
frame of the current paradigm a new family plays a key role, namely x1mul2,
which is of multiplicity two. This family of p.o. can either reinforce b/p
side-on profiles based mainly on a x1v1-x1v1$^{\prime}$ backbone, or even
support a b/p morphology just by itself. In more detail our conclusions can be
summarised as follows:

\begin{enumerate}
 \item We find non-periodic orbits supporting a b/p profile in the side-on view
\textit{before} the critical energy in the v-ILR region at which x1v1 is
bifurcated from x1. These orbits are quasi-periodic orbits on tori around stable
x1 orbits. Their outline tends to a combined x1v1-x1v1$^{\prime}$ morphology as
\ej increases towards the value at which x1v1 is introduced in the system
(Fig.~\ref{thickx1}). This tells us firstly that the peanut starts before the
v-ILR resonance. In our model, peanut-supporting orbits are found
extending along  the major axis of the bar already at y=0.13 bar lengths. 
Secondly, that the structure of the phase space in the
neighbourhood of the mother family changes in a way that the coming of the
bifurcated family can be foreseen by the gradual change in the shape of its
quasi-periodic orbits. We do not have an abrupt change of the phase space
structure immediately upon the introduction of the bifurcated family in the
system. 
  \item Non-periodic orbits in the neighbourhood of x1 reinforce boxy profiles
even in the interval in which x1 is simple unstable. These are orbits, which
either become sticky to the nearby x1v1 and x1v1$^{\prime}$ rotational tori, or
in general orbits that can be observed being trapped in cylindrical structures
in 3D projections of the 4D spaces of section that include the $(x,p_x)$ plane.
This is a significant dynamical mechanism for building the body of the
peanut.
  \item The neighbourhood of x1 p.o. is a substantial
source of non-periodic orbits that contribute to boxy side-on profiles. They
provide such orbits to the system before the v-ILR and in the $\Delta
$E$_{vILR}$ region where it is U. Of particular significance is its
contribution in regions where x1 is the only stable family of p.o. in the
system (in our model $-0.332<$\ej$<-0.302$). These orbits contribute to a
b/p morphology mainly by becoming sticky to x1v1 and x1v1$^{\prime}$
rotational tori. Else, they contribute to an overall edge-on boxiness.
  \item In the $\Delta$E$_{vILR}$ region, apart from the U x1 p.o. we have also
the U x3 p.o. The structure of the phase space around them is different, since
they are unstable to vertical and radial perturbations respectively. The
structure of the phase space in the neighbourhood of U p.o. is determined by
the type of simple instability of the p.o.
  \item Edge-on profiles of non-periodic orbits in the neighbourhood of x2 and
x3 have a certain similarity (Fig.~\ref{x1x3per}). Thus they will have similar
contribution to b/p profiles of galactic bars. They reach heights about 100~pc,
considerably smaller than the heights of the boxy bulges in images of edge-on
galaxies. This indicates that if x2 stellar flows exist in 3D stellar bars, they
are embedded inside the edge-on profiles of the b/p structures.
  \item In the case of {\sf X}-shaped features built by orbits associated
with the x1v1 family, the ``{\sf X}'' is formed by the tips of orbits with
rather ``$\smile$''-like or ``$\frown$''-like morphologies in their side-on
views.
Along the wings of ``{\sf X}'' we have $p_z=0$ (Fig.~\ref{noperorbs}).
The morphology of a b/p structure is always characterised by the outermost
orbits. Models in which we observe sharp wings emerging out of boxy structures
can be constructed by populating orbits in the x1v1 neighbourhood at \ej beyond
the $\Delta$ part of the x1v1 family. The part of the b/p bulge
supported by chaotic orbits at energies where x1v1 is $\Delta$, is expected to 
have less conspicuous ``{\sf X} - wings''. Bars with moderate amplitudes can 
support more efficiently a {\sf X}-shaped b/p bulge based on x1v1
dynamics, because in this case this family is complex unstable over smaller
$\Delta$E$_{J}$ ranges.
  \item ``CX-type'' profiles can be supported:
\vspace{-1mm}
\begin{enumerate}
  \item By quasi-periodic orbits around x1 in energies, where the x1v2 family
    co-exists. They become more important as we increase their $p_{z_0}$
    initial condition from 0 towards the $p_{z_0}$ value of x1v2. The strongest
    reinforcement is given by orbits corresponding to the closest to x1v2 tori
    around x1. 
  \item By sticky chaotic orbits around the x1v1 and x1v1$^{\prime}$ tori. We
    find such orbits by perturbing the x1v2 initial conditions, since the
    unstable asymptotic curves around x1v2 wind around the tori of the x1v1 and
    x1v1$^{\prime}$ families.
\end{enumerate}
 \vspace{-1mm}
 \item A very important role for the dynamics of the b/p bulge is played by the
family of multiplicity 2, x1mul2. This family exists already at energies
smaller than the one at which x1v1 is introduced in the system, being stable
over large $\Delta$\ej energy intervals. Quasi-periodic orbits trapped around
the orbits of this family and, mainly, sticky orbits to them reinforce a b/p
structure.  
 \item We find chaotic orbits in the neighbourhood of complex unstable
periodic orbits that can contribute to the reinforcement of a b/p structure for
several tens of dynamical times. In our model, in the case of chaotic orbits
close to complex unstable x1v1 p.o. there is a correlation between the degree of
confinement of the chaotic orbit in the configuration space and the value of
the quantity $\Delta$ (Eq.~\ref{eq:delta}). As we have seen in
Fig.~\ref{noperorbs}, the smaller the quantity $\Delta$ is, the largest the
deviation of its morphology from that of a quasi-periodic orbit.
Interesting for the general theory of
Hamiltonian systems is the correspondence found between the structure of
the phase space in the neighbourhood of simple and double complex unstable p.o.
In the case of p.o. of double multiplicity the 4D spiral structure is
developed in two confined tori simultaneously. We see also that tiny
perturbations of complex unstable periodic orbits do not lead to an abrupt
filling of the available phase space by the consequents of the two orbits. In a
case with two ``$\Delta$'' p.o. (one of simple and one of double
multiplicity), which co-exist at the same \ej, we have found orbits in their
immediate neighbourhood that stay in different regions of the phase space for
thousands of consequents. 
  \item Considering non-periodic orbits at successive energies introduced in the
system through one of the dynamical mechanisms described in our paper, one can
build conspicuous b/p profiles mainly in their side-on views. Away from the
side-on view we have in general just a boxy morphology in the outer envelopes
of the edge-on projections. However, in the case of {\sf X}-shaped boxy bulges
based on x1v1
dynamics, we find nearly end-on views, where the same orbits can give an
``$\infty$''-type morphology. 
\end{enumerate}

\vspace{0.5cm} 
\noindent \textit{Acknowledgements}

We thank Prof. G.~Contopoulos for fruitful discussions and very useful comments.
This work has
been partially supported by the Research Committee of the Academy of
Athens through the project 200/815.

\label{lastpage}


\begin{thebibliography}{}
\bibitem[Aronica (2006)]{ar06} Aronica G., 2006, PhD Thesis,
Ruhr-Universit\"{a}t Bochum
\bibitem[Athanassoula \& Misiriotis (2002)]{am02} Athanassoula E., Misiriotis
A., 2002, MNRAS 330, 35
\bibitem[Broucke (1969)]{br69} Broucke R., 1969, ``Periodic orbits in the
  elliptic restricted three-body problem'', NASA Techn. Rep. 32, 1360
\bibitem[Bountis et al. (2012)]{bma}Bountis T., Manos T., Antonopoulos C.,
2012, Cel.Mech.Dyn.Ast. 113, 63
\bibitem[Bureau \& Athanassoula (2005)]{ba05} Bureau M., Athanassoula
E., 2005, ApJ 626, 159
\bibitem[Bureau et al. (2006)]{betal06} Bureau M., Aronica G., Athanassoula
E., Dettmar R.-J., Bosma A., Freeman, K. C., 2006, MNRAS, 370, 753
\bibitem[Cheung et al. (2013)]{chetal13} Cheung E., Athanassoula E., Masters
K.L. et al. 2013, ApJ 779, 162
\bibitem[Combes et al. (1990)]{cetal90} Combes F., Debbasch F., Friedli D.,
Pfenniger D., 1990, A\&A 233, 95  
\bibitem[Contopoulos (1981)]{gco81} Contopoulos G., 1981, A\&A 102, 265
\bibitem[Contopoulos (2004)]{gcobook} Contopoulos G., 2004, ``Order and Chaos in
Dynamical Astronomy'', Springer Verlag, Berlin-Heidelberg
\bibitem[Contopoulos \& Barbanis (1985)]{cb85} Contopoulos G., Barbanis B.,
1985, A\&A 153, 44
\bibitem[Contopoulos \& Grosb{\o}l (1989)]{cg89} Contopoulos G., Grosb{\o}l
P., 1989, A\&AR, 1,261
\bibitem[Contopoulos \& Harsoula (2010)]{ch10} Contopoulos G., Harsoula M.,
2010, Cel.Mech.Dyn.Ast. 107, 77
\bibitem[Contopoulos \& Harsoula (2013)]{ch13} Contopoulos G., Harsoula M.,
2013, MNRAS 436, 1201
\bibitem[Contopoulos \& Magnenat (1985)]{cm85} Contopoulos G., Magnenat P.,
1985, Celest. Mech. 37, 387
\bibitem[Contopoulos et al. (1994)]{cfpp94} Contopoulos G., Farantos S.C.,
Papadaki H., Polymilis C., 1994, Cel. Mech. 37, 387
\bibitem[Erwin (2005)]{e05} Erwin P., 2005, MNRAS 364,283
\bibitem[Gardner et al. (2014)]{eskoetal14}Gardner E., Debattista V., Robin A.,
et al. 2014, MNRAS 438, 3275
\bibitem[Hadjidemetriou (1975)]{h75} Hadjidemetriou J., 1975, Celest. Mech. 12, 
255  
\bibitem[Heggie (1985)]{dh85} Heggie D.C., 1985, Celest. Mech. 35, 357
\bibitem[Hickson (1993)]{hcks93} Hickson P., 1993, Astr. Lett. Comm. 29, 1
\bibitem[Jorba \& Olle (2004)]{jo04} Jorba A., Olle M., 2004, 
  Nonlinearity 17, 691
\bibitem[Katsanikas \& Patsis (2011)]{kp11} Katsanikas M., Patsis P.A., 2011, 
  Int. J. Bif. Ch. 21-02, 467
\bibitem[Katsanikas et al. (2011a)]{kpc11} Katsanikas M., Patsis P.A.,
Contopoulos G., 2011a, Int. J. Bif. Ch. 21-08, 2321
\bibitem[Katsanikas et al. (2011b)]{kpp11} Katsanikas M., Patsis P.A.,
Pinotsis A.D., 2011b, Int. J. Bif. Ch. 21-08, 2331
\bibitem[Katsanikas et al. (2013)]{kpc13} Katsanikas M., Patsis P.A.,
Contopoulos G., 2013, Int. J. Bif. Ch. 23-02, 1330005
\bibitem[Kregel et al. (2004)]{kvkf04} Kregel M., van der Kruit P.C.,
Freeman K.C., 2004, MNRAS 351, 1247
\bibitem[Lange et al. (2014)]{langetal} Lange S., Richter M.,
Onken F. et al., 2014, arXiv:1311.7632 [nlin.CD]
\bibitem[Li et al. (2012)]{ls12} Li Z-Y., Shen J. 2012, ApJL 757L, L7
\bibitem[L\"{u}tticke et al. (2000)]{ldp00} L\"{u}tticke R., Dettmar R.-J.,
Pohlen M., 2000, A\&A 362, 435
\bibitem[Martinez-Valpuesta et al. (2006)]{m-vetal06} Martinez-Valpuesta I.,
Shlosman I., Heller C., 2006, ApJ 637, 214
\bibitem[Miyamoto \& Nagai (1975)]{miyna75} Miyamoto M., Nagai R., 1975, PASJ,
27,533.
\bibitem[Olle et al. (2000)]{oea04} Olle M., Pacha J.R.,
Villanueva J., 2004, Cel. Mech. Dyn. Astr. 90, 89
\bibitem[Papadaki et al. (1995)]{pcp95} Papadaki H., Contopoulos G., Polymilis
C., 1995, in \textit{``From Newton to Chaos'}', Roy A.E., Steves B.A. (eds), pp
485-494, Plenum Press, New York
\bibitem[Patsis (2005)]{p05} Patsis P.A., 2005, MNRAS 358, 305
\bibitem[Patsis \& Grosb{\o}l (1996)]{pg96} Patsis P.A., Grosb{\o}l
P., 1996, A\&A 315, 371
\bibitem[Patsis \& Katsanikas (2014)]{pII} Patsis P.A., Katsanikas M., 2014,
MNRAS (in press).

\bibitem[Patsis \& Xilouris (2006)]{px06}  Patsis P.A., Xilouris E., 2006,
MNRAS 366, 1121
\bibitem[Patsis \& Zachilas (1990)]{pz90}  Patsis P.A., Zachilas L., 1990, 
A\&A 227, 37
\bibitem[Patsis \& Zachilas (1994)]{pz94}  Patsis P.A., Zachilas L., 1994, Int.
J. Bif. Ch. 4, 1399 
\bibitem[Patsis et al. (2002)]{psa02} Patsis P.A., Skokos Ch., Athanassoula E.,
2002, MNRAS 337, 578
\bibitem[Pfenniger (1984)]{pf84} Pfenniger D., 1984, A\&A 134, 373
\bibitem[Pfenniger (1985a)]{pf85a} Pfenniger D., 1985a, A\&A 150, 97
\bibitem[Pfenniger (1985b)]{pf85b} Pfenniger D., 1985b, A\&A 150, 112  
\bibitem[Plummer (1911)]{pl11} Plummer H.C., 1911, MNRAS 71, 460
\bibitem[Quillen et al. (2014)]{qetal14} Quillen A.C., Minchev I., Sharma S. et
al., 2014, MNRAS 437, 1284
\bibitem[Richter et al. (2013)] {retal13} Richter M., Lange S.,  B\"{a}cker A., 
Ketzmerick R., 2014, PhRvE, 89,022902
\bibitem[Saha \& Gerhard (2013)]{so13} Saha K., Gerhard O., 2013, MNRAS 430,
2039
\bibitem[Saha \& Naab (2013)]{sn13} Saha K., Naab T., 2013, MNRAS 434, 1287
\bibitem[Saito et al (2011)]{szw11} Saito R.K., Zoccali M., Mc William A. et
al., 2011, AJ 142, 76
\bibitem[Skokos (2001)]{sk01} Skokos Ch., 2001, Physica D 159, 155
\bibitem[Skokos et al. (2002a)]{spa02a}  Skokos Ch., Patsis P.A., Athanassoula
   E., 2002a, MNRAS 333, 847
\bibitem[Skokos et al. (2002b)]{spa02b}  Skokos Ch., Patsis P.A., Athanassoula
   E., 2002b, MNRAS 333, 861
\bibitem[V\'{a}squez et al. (2013)]{vzhetal13} V\'{a}squez S., Zoccali M., Hill
V. et al. 2014, A\&A 555, 91
\bibitem[Vrahatis et al. (1997)]{vetal97} Vrahatis M., Isliker H., Bountis
T.C., 1997, Int. J. Bif. Ch. 7, 2707
\bibitem[Wakamatsu \& Hamabe (1984)]{wh84} Wakamatsu K., Hamabe M., 1984, ApJS
56, 283
\bibitem[Wegg \& Gerhard (2013)]{wg13} Wegg C., Gerhard O., 2013, MNRAS 435,
1874
\bibitem[Williams et al. (2011)]{wetal11} Williams M. J, Zamojski M. A.,
Bureau M. et al., 2011, MNRAS 414, 2163
\bibitem[Zachilas (1993)]{z93} Zachilas L., A\&AS 97, 549
\bibitem[Zachilas et al. (2013)]{zkp13} Zachilas L.,  Katsanikas M., Patsis P.
A. 2013, Int. J. Bif. Ch. 23, 1330023 
\end{thebibliography}
\end{document}